\documentclass{article}
\usepackage[utf8]{inputenc}

\newsavebox{\foobox}
\newcommand{\slantbox}[2][0]{\mbox{%
        \sbox{\foobox}{#2}%
        \hskip\wd\foobox
        \pdfsave
        \pdfsetmatrix{1 0 #1 1}%
        \llap{\usebox{\foobox}}%
        \pdfrestore
}}
\newcommand\unslant[2][-.25]{\slantbox[#1]{$#2$}}

\newcommand{\mdelta}{\text{\unslant[-.18]\delta}}

\usepackage[left=2.5cm,right=2.5cm, top=1in, bottom=1in]{geometry}

\usepackage{isomath}
\usepackage{euscript}
\usepackage{amssymb}
\usepackage{amsfonts}
\usepackage{amsmath}
\usepackage{amsbsy}
\usepackage{epsfig}
\usepackage{amsthm}
\usepackage{amscd}
\usepackage{amstext}
\usepackage{verbatim}
\usepackage{amsmath}
\usepackage{cancel}
\usepackage{capt-of}
\usepackage{empheq}
\usepackage{slashed}
\usepackage{xcolor}
\usepackage{setspace}
\usepackage{dsfont}

\usepackage[colorlinks=true, urlcolor=violet, linkcolor=blue, citecolor=red, hyperindex=true, linktocpage=true]{hyperref}

\newcommand{\red}[1]{{\color{black} #1\color{black}}}
\newcommand{\blue}[1]{{\color{black} #1\color{black}}}

\def\ben{\begin{equation}}
\def\een{\end{equation}}

\def\be{\begin{equation}}
\def\ee{\end{equation}}
\def\beq{\begin{equation}}
\def\eeq{\end{equation}}
\def\ba{\begin{array}}
\def\ea{\end{array}}

\def\dalemb#1#2{{\vbox{\hrule height .#2pt
       \hbox{\vrule width.#2pt height#1pt \kern#1pt
               \vrule width.#2pt}
       \hrule height.#2pt}}}

\newcommand{\bea}{\begin{eqnarray}}
\newcommand{\eea}{\end{eqnarray}}

  \usepackage{titlesec}

\renewcommand\theparagraph{\thesubsubsection.\arabic{paragraph}}

\makeatletter

\DeclareRobustCommand\bfseriesitshape{%
  \not@math@alphabet\itshapebfseries\relax
  \fontseries\bfdefault
  \fontshape\itdefault
  \selectfont
}

\titleformat{\paragraph}
{\normalfont\normalsize\bfseries}{\theparagraph}{1em}{}
\titlespacing*{\paragraph}
{0pt}{3.25ex plus 1ex minus .2ex}{1.5ex plus .2ex}
\def\toclevel@paragraph{4}
\def\l@paragraph{\@dottedtocline{4}{7em}{4em}}
\makeatother

\begin{document}

\allowdisplaybreaks

\thispagestyle{empty}

\begin{center}

{ \Large {\bf
Kinetic theory of transport for inhomogeneous electron fluids
}}

\vspace{1cm}

Andrew Lucas and Sean A. Hartnoll

\vspace{1cm}

{\small{\it 
 Department of Physics, Stanford University, \\
Stanford, CA 94305-4060, USA \\
\vspace{0.3cm}

 }}

\vspace{1.6cm}

\end{center}

\begin{abstract}
The interplay between electronic interactions and disorder is neglected in the conventional Boltzmann theory of transport, yet can play an essential role in determining the resistivity of unconventional metals.  When quasiparticles are long-lived, one can account for these intertwined effects by solving spatially inhomogeneous Boltzmann equations.  Assuming smooth disorder and neglecting umklapp scattering,  we solve these inhomogeneous kinetic equations and compute the electrical resistivity across the ballistic-to-hydrodynamic transition. 
An important consequence of electron-electron interactions is the modification of the momentum relaxation time; this effect is ignored in the \red{homogeneous} theory.   We
characterize precisely when interactions enhance the momentum scattering rate, and when they decrease it. Our approach unifies existing semiclassical theories of transport  and reveals novel transport mechanisms.  In particular, we explain how the resistivity can be proportional to the rate of momentum-conserving collisions.  We compare this result with existing transport mysteries, including the disorder-independent $T^2$ resistivity of many Fermi liquids, and the linear-in-$T$ ``Planckian-limited" resistivity of many strange metals.
\end{abstract}

\vspace{6.6cm}

\noindent \texttt{ajlucas@stanford.edu}
\\
\texttt{hartnoll@stanford.edu}

\pagebreak
\setcounter{page}{1}

\tableofcontents

\section{Introduction}

\subsection{The Challenge of Metallic Transport}
\label{sec:intro1}
One of the simplest experiments a condensed matter physicist can perform is to measure the electrical resistivity $\rho$ of a metal.  Unfortunately, computing the resistivity from first principles is extremely challenging. 
This is because the resistivity crucially depends on (\emph{i}) the scattering rates and pathways of the electrons and (\emph{ii}) the mechanism through which translation invariance is lost.  

To understand the heart of the challenge, let us briefly review the origins of transport theory.   In many of the most common metals, a
simple picture proposed by Drude \cite{drude} in 1900 holds quite well.  We estimate that the resistivity \begin{equation}
\rho = \frac{m}{ne^2} \frac{1}{\tau},  \label{eq:drude}
\end{equation}
with $m$ the effective mass of quasiparticles in the metal, $n$ the density of quasiparticles, and $\tau^{-1}$ is the `scattering rate' of these quasiparticles.  Bloch \cite{Bloch1, Bloch2} improved on this picture in 1929, noting that $\tau^{-1}$ ought to be the rate at which quasiparticles lose their momentum.   However, it was already appreciated by Peierls \cite{Peierls1, Peierls2} in 1930  that such a picture has a serious caveat:  whatever the quasiparticles scatter off of must rapidly relax the \emph{total} momentum of the system.   There is a simple argument:  if the total momentum of the system is conserved, then we may shift to a reference frame moving at velocity $\vec v$ relative to the crystal rest frame.  In the new reference frame, we observe an electric current $\vec J = -en \vec v$.  However, Peierls' critique turns out to be unimportant for common metals, where most scattering events can relax momentum.  

For these common metals, the theory of transport was placed on solid ground 60 years ago, e.g.  \cite{ziman}.  One calculates the rate at which quasiparticles of momentum $\vec p$ scatter into quasiparticles of momentum $\vec q$.  By associating this with the collision integral of a homogeneous, linearized Boltzmann equation, one is easily able to compute the resistivity of a metal.  Bloch's and Peierls' improvements have been accounted for. 

However, we have now seen many materials whose transport properties are still beyond the conventional paradigm.  A well-known failure of textbook theory arises in ``strange metals" where one commonly measures $\rho \propto T$ at temperatures well below the Debye temperature.  Upon closer analysis, one finds \cite{andy} \begin{equation}
    \rho \approx \rho_0 +  \frac{m}{ne^2}\frac{k_{\mathrm{B}}T}{\hbar}. \label{eq:rholinear}
\end{equation}
  The linear in $T$ contribution to $\rho$ is consistent with the Drude formula (\ref{eq:drude}) if there is a scattering rate \begin{equation}
      \frac{1}{\tau} \approx \frac{k_{\mathrm{B}}T}{\hbar}. \label{eq:Trate}
  \end{equation} This is precisely the scattering rate of a strongly interacting, quantum critical strange metal. It may also be the
``fastest scattering rate" in nature for \emph{momentum-conserving} collisions, that lead to the loss of quantum coherence \cite{ssbook, Maldacena:2015waa, Hartnoll:2016apf}. One of us \cite{Hartnoll:2014lpa} conjectured that (\ref{eq:rholinear}) may arise from saturating a fundamental bound on transport, where momentum relaxing collisions also occur at the rate (\ref{eq:Trate}).  However, an immediate problem with bounding the rate of momentum relaxing collisions as $\propto T$, even in a metallic state, arises from the fact that the constant $\rho_0$ in (\ref{eq:rholinear}) is widely believed to arise from scattering off of static impurities. \red{Indeed $\rho_0$ is strongly disorder-dependent while the coefficient of the $T$-linear term is not \cite{PhysRevLett.76.684}.}  If the universality of (\ref{eq:rholinear}) arises from the universality of (\ref{eq:Trate}), then physics beyond the Drude paradigm must be responsible.
  
A less well-appreciated failure of the textbook theory arises in the conventional Fermi liquid phase of many `complicated' metals, including heavy fermion metals.   Here one measures the resistivity \begin{equation}
    \rho = \rho_0 + AT^2,  \label{eq:AT2}
\end{equation}
where the coefficient $A$ typically depends most strongly on the thermodynamic properties of the sample.  In fact, there appear to be universal relationships between $A$ and simple thermodynamic properties such as the specific heat and the band structure \cite{KADOWAKI1986507, Jacko2009}.  Again, it appears that translation symmetry breaking plays no role in determining the coefficient $A$ arising in the resistivity, which is at odds with the theorem that $\rho>0$ is solely a consequence of translation symmetry breaking.    So we have a second example where, neglecting the constant $\rho_0$,  the resistivity seems directly tied to a scattering rate most easily associated with momentum-conserving collisions.  Given a diverse array of sample quality and material structure, a theory which leads to (\ref{eq:AT2}), where disorder plays a minimal role in determining the coefficient $A$, is clearly needed.

A common explanation for the $T^2$ resistivity of the heavy fermion and other materials is that (\emph{i}) there are multiple bands present, of different quasiparticle masses \cite{Baber383}, and/or (\emph{ii}) that typical electron-electron scattering is an umklapp process, which can directly relax momentum \cite{ziman}.   The first explanation requires that a ``heavy" band efficiently relax momentum \red{or that the metal has perfectly compensated electron and hole Fermi surfaces so that the total charge density is zero}.  The second explanation is plausible so long as the band structure permits efficient umklapp scattering near the Fermi surface.   However, the universality of $A$ renders this proposal rather unappealing, given the diverse band structures present in these different compounds, which ought to lead to differences in the efficiency of umklapp between different materials.  We also note that recent experiments on the $\mathrm{SrTiO}_3$ also show anomalous $T^2$ resistivity: in the material there is only a single band of electrons at the Fermi surface, and umklapp is highly suppressed \cite{kamran, suzanne}.

\subsection{Kinetic Theory Beyond the Relaxation Time Approximation}
\label{sec:kineticintro}
\blue{The conventional theory of transport in condensed matter physics \cite{ziman} -- which is based on kinetic theory} -- is not sophisticated enough to solve these puzzles. 
This is of course true for the `most strongly correlated' metals, where no quasiparticles exist -- the key assumption underpinning the kinetic equations is the existence of quasiparticles.  However, there are many metals where (\emph{i}) quasiparticle-quasiparticle scattering is important and (\emph{ii}) quasiparticles remain long-lived.  This occurs whenever the electron-electron mean free path is shorter than the electron-impurity mean free path.   Such a regime has been accessed in experiments on multiple materials \cite{Molenkamp95, Bandurin1055, Moll1061, 2017arXiv170306672K,WP2}. While we will often refer to  `electron-electron' scattering in this work, we technically always mean `quasiparticle-quasiparticle' scattering.   Transport in these metals, with long-lived quasiparticles, is still beyond the conventional framework, as we now explain.
  
The reason that the textbook kinetic theory of transport \cite{ziman} is not suitable for such systems is that it neglects classical correlations between scattering events.   The approximation that particles are equally likely to scatter from momentum $\vec p$ into $\vec q$, everywhere in the sample, is simply not true in general.  For example, consider a quasiparticle moving through a slowly varying potential.  If this quasiparticle collides many times with other quasiparticles in one `patch' of the potential, then the local transition rates are \emph{not} equivalent to spatially averaged transition rates.  
\red{In this limit, one can model transport using hydrodynamics, as has been done in older \cite{Gurzhi63, Gurzhi68} as well as more recent work \cite{KS06, KS11, Davison:2013txa, Lucas:2015lna, PhysRevB.93.075426, ushydro}.}

Kinetic theory was invented to study the dynamics of gases and to compute the viscosity of air.  Because hydrodynamics is contained as a special limit of kinetic theory, a correct and complete solution of the kinetic theory of transport must recover the hydrodynamic limit of transport when the electron-electron mean free path is sufficiently short.
\red{Previously, the homogeneous Boltzmann equation has been solved in finite geometries, where boundary conditions play the role of `disorder' and lead to non-trivial transport phenomena acrosss the ballistic-to-hydrodynamic crossover \cite{Molenkamp95, levitov2,2016arXiv161200856L,levitov3}.    There is also previous literature on transport of non-interacting electrons in disorder potentials by perturbatively solving the Liouville (non-interacting Boltzmann) equation \cite{mirlin}. Our work will borrow some techniques from these works and extend them into new regimes:  we directly account for both the disorder inevitably present in the bulk of the sample and the effects of electron-electron interactions.}

\red{In this paper, we solve the kinetic theory of transport in an inhomogeneous system. We are able to compute the resistivity of a disordered medium across the ballistic-to-hydrodynamic crossover, recover all known (semi)classical transport phenomena within a unified framework and identify new hydrodynamic regimes. We answer the question: when do interactions enhance the momentum scattering rate, and when do they decrease it? We present two explicit types of calculations.  When the inhomogeneity is weak, we integrate it out and exactly compute the resistivity to leading order in perturbation theory.   When the inhomogeneity is strong, we present a variational principle for upper bounding the resistivity.   Both techniques are completely general and valid for any system with long lived quasiparticles and inversion and time-reversal symmetry. The techniques can immediately be applied to realistic, material-specific models of electronic transport.}

\section{Summary of Results}
\label{sec:summary}
We first introduce the model and explain qualitatively the transport phenomena that are possible.   We then describe the predictions of our formalism for experiments, and comment on the similarities between our findings and the experimental mysteries we outlined in Section \ref{sec:intro1}.

\subsection{Transport Regimes}

In this paper, we will consider a toy model for kinetic transport which elucidates the failures of the \red{relaxation time approximation}.   We consider a weakly interacting gas of long-lived quasiparticles of Fermi wavelength $\lambda_{\mathrm{F}}$, moving through a smooth disorder potential $V_{\mathrm{imp}}(\vec x)$,  \begin{figure}[h]
\centering
\includegraphics[width=5in]{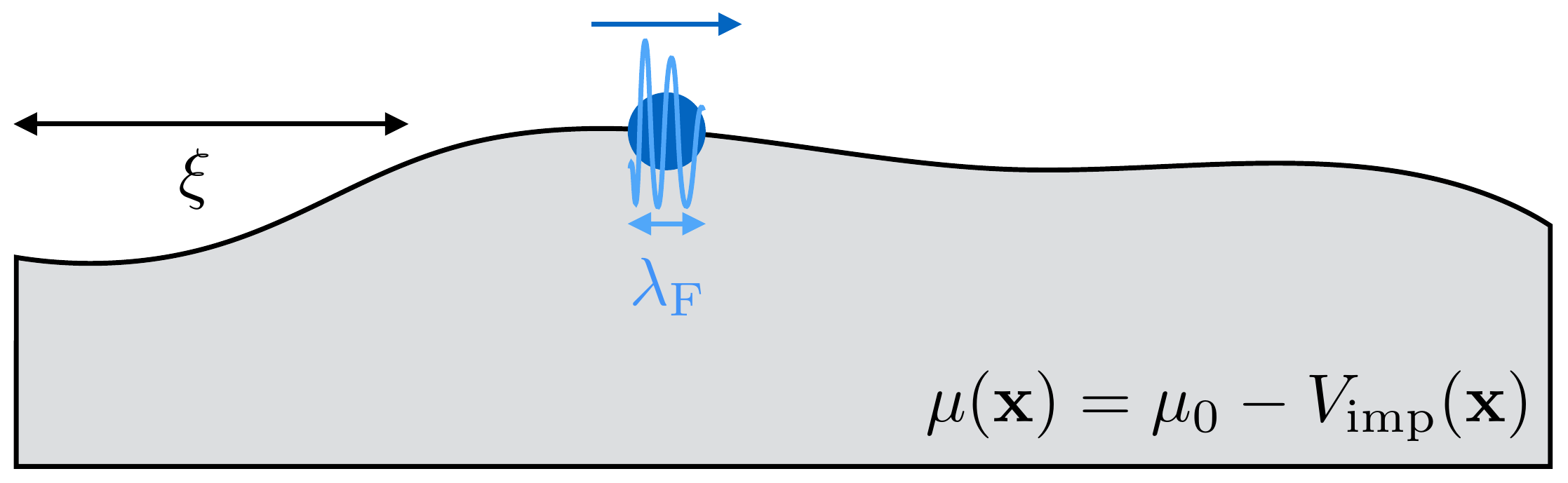}
\caption{Quasiparticles of short Fermi wavelength $\lambda_{\mathrm{F}}$ moving in a smooth disorder potential $V_{\mathrm{imp}}$.  In the many-body limit, this disorder potential may be interpreted as an inhomogeneous chemical potential.}
\label{fig:vimp}
\end{figure}
which varies on the length scale $\xi$: see Figure \ref{fig:vimp}. 
The Hamiltonian of the many-body system is \begin{equation}
    H = \sum_{i=1}^N \left(\epsilon(\vec p_i) + V_{\mathrm{imp}}(\vec x_i)\right) + H_{\mathrm{int}},
\end{equation}
where $\epsilon(\vec p) = \epsilon(-\vec p)$ is an inversion-symmetric kinetic energy (which need not be $\vec p^2/2m$) and $H_{\mathrm{int}}$ is a many-body Hamiltonian allowing for interactions between the quasiparticles.   We assume that $H_{\mathrm{int}}$ is invariant under uniform translations, and so the total momentum of the electrons is conserved in the absence of the impurity potential.  When $\lambda_{\mathrm{F}} \ll \xi$, we will show in Section \ref{sec:boltz} that one must account for $V_{\mathrm{imp}}$ in a more sophisticated manner than is conventionally done \cite{ziman}, and solve the spatially inhomogeneous kinetic equations.  While our focus in this paper is on the limit where momentum relaxation is entirely due to long wavelength disorder,  it is straightforward to add short-range impurity scattering, umklapp and/or phonon scattering.  These will add conventional momentum-relaxing contributions to the kinetic equations.

We will present two techniques for solving the inhomogeneous kinetic equations.   Firstly in Section \ref{sec:perturb}, when $V_{\mathrm{imp}}$ is weak, we exactly integrate it out and compute $\rho$ at leading order.   Secondly, in Section \ref{sec:bounds}, we prove a variational principle which can be used to compute upper bounds on the resistivity even when the inhomogeneity is non-perturbatively large.   Schematically, our variational principle gives that \begin{equation}
    \rho_{xx} \le  \left.\frac{\frac{1}{V}\int \mathrm{d}^dx \, T\dot{s}}{(\frac{1}{V} \int \mathrm{d}^x \, J_x)^2}\right|_{\nabla \cdot J^A=0};  \label{eq:entcart}
\end{equation} 
the resistivity can be computed by minimizing the entropy produced on \emph{arbitrary} small deviations away from equilibrium,  subject to the constraint that all conservation laws (including charge) are respected:  $\nabla \cdot J^A=0$. \red{There are integrals over space (normalized by the total volume $V$), but no disorder average in (\ref{eq:entcart}). In particular, the constraint must be obeyed in a specific inhomogeneous potential. This constraint is the new ingredient relative to older variational principles for the homogeneous Boltzmann equation \cite{Kohler1948,Kohler1949,Sondheimer75,zimanshort,ziman}.} As explicit applications of these techniques, we have studied models where the impurity potential is characterized by a single length scale $\xi$,  where electron-electron collisions occur on a fixed length scale $\ell_{\mathrm{ee}}$, and where all thermally excited quasiparticles move at a typical speed $v_{\mathrm{F}}$.   Our formalism is applicable for more complicated systems, though we leave detailed analyses of these generalizations to future work.  Both the exact perturbative and variational non-perturbative computations suggest that there are three main regimes of transport, summarized in Figure \ref{fig:cartoon}:

\begin{figure}[h]
\centering
\includegraphics[width=6.4in]{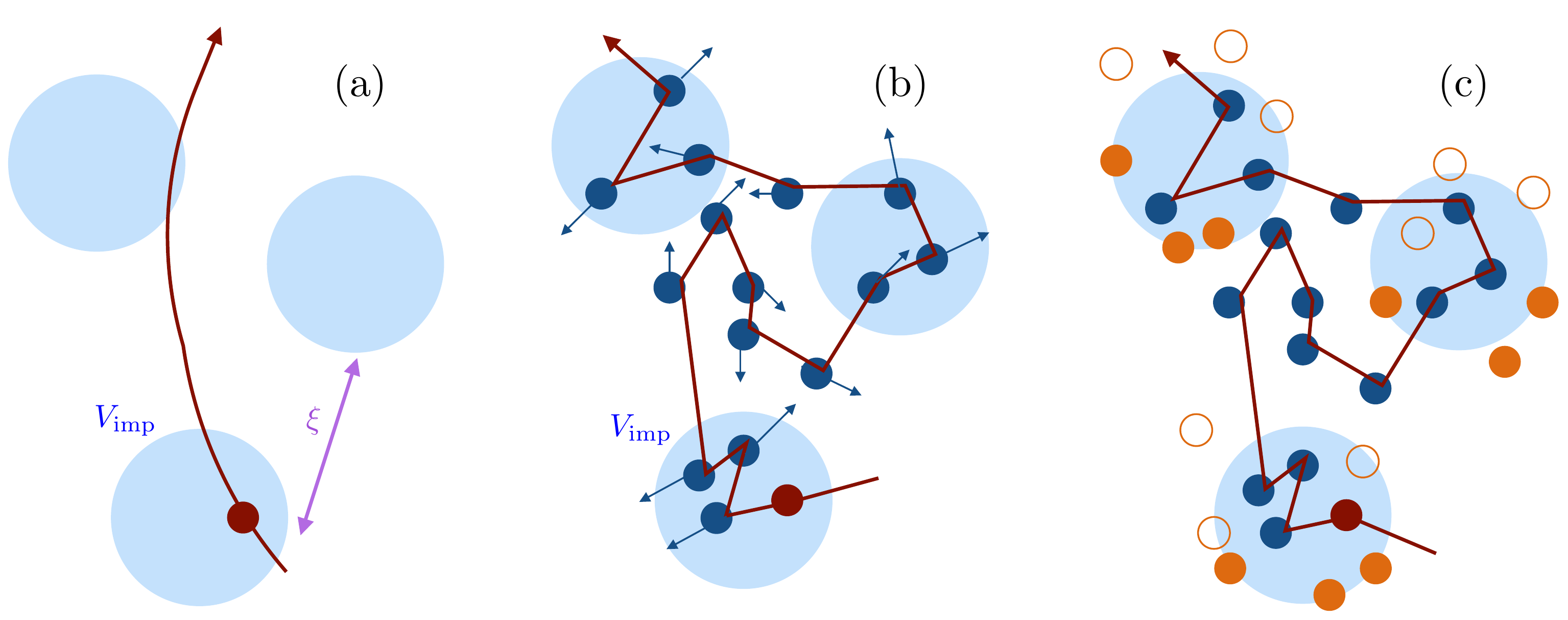}
\caption{We follow a ``special" quasiparticle (in red) as it meanders through a disordered landscape, possibly colliding with other quasiparticles.  In each plot, we show a birds eye view of a red quasiparticle, moving through the inhomogeneous impurity potential $V_{\mathrm{imp}}$ (shown in light blue) of a two-dimensional Fermi liquid. This is an artistic simplification;  our results are not specific to $d=2$.  (a) A non-interacting quasiparticle random walks through the impurity potential, deflected by a small angle after each puddle.  This leads to (\ref{eq:mainbal}).  (b) A quasiparticle rapidly collides with others (in blue) on a length scale $\ell_{\mathrm{ee}}$.  The time it takes for each quasiparticle to ``see" the inhomogeneity is thus enhanced, leading to (\ref{eq:mainvisc}).  The arrows on each blue particle emphasize that these collisions conserve the total momentum, and cannot directly contribute to the resistivity. (c) A large number of conservation laws ensure that as we try to move any one quasiparticle (in red) across the puddle, many other quasiparticles are forced out of equilibrium in order to satisfy additional conservation laws:  the orange holes denote the `absence' of quasiparticles, out of equilibrium, and the solid orange circles denote where the quasiparticles have been driven.  The large number of quasiparticles out of equilibrium leads to the effective momentum relaxation rate (\ref{eq:mainimb}).}
\label{fig:cartoon}
\end{figure}

\begin{figure}
\centering
\includegraphics[width=3.3in]{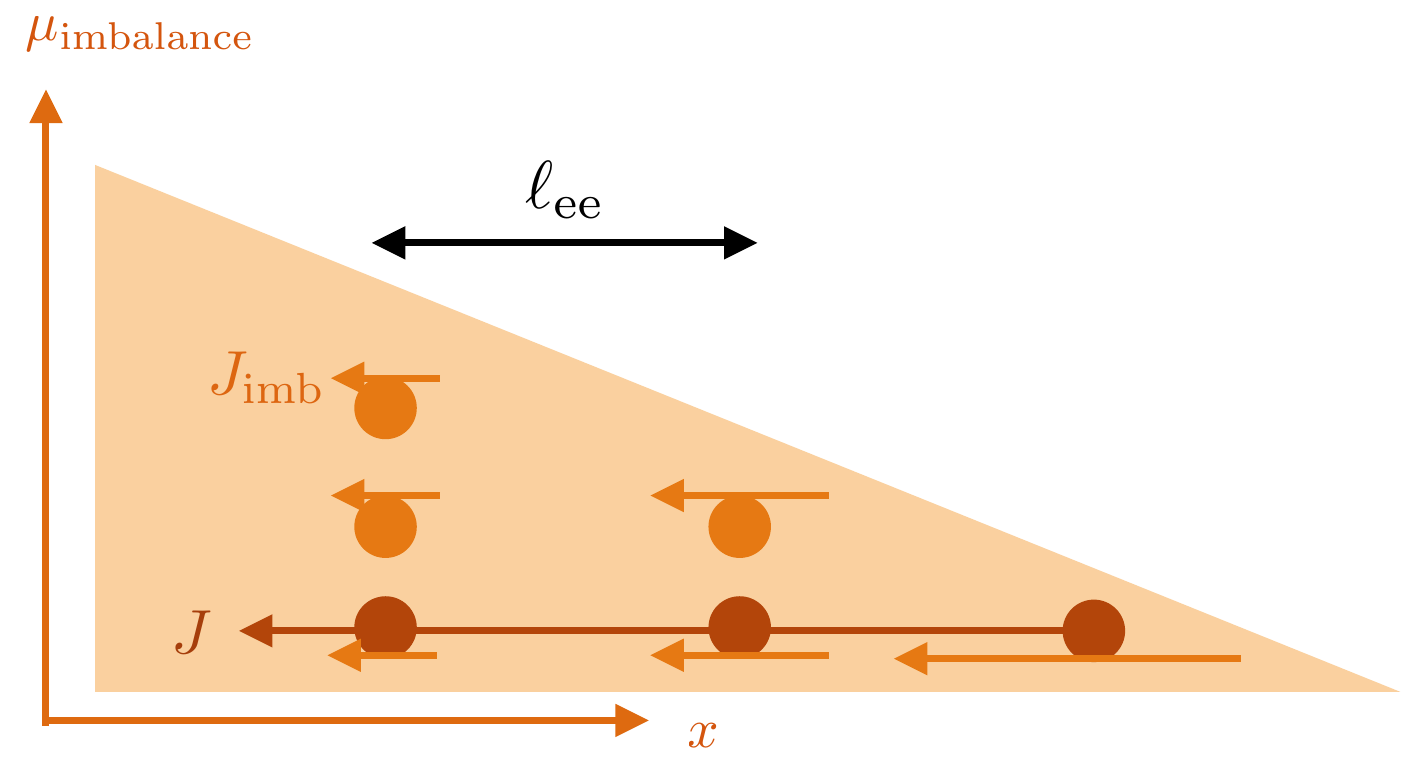}
\caption{In order to conserve all currents as we move through the inhomogeneous landscape, we must excite $N_{\mathrm{imbalance}}$ quasiparticles out of equilibrium.}
\label{fig:imbfig}
\end{figure}

\begin{enumerate}

\item \textbf{ballistic:}  In the limit where $\ell_{\mathrm{ee}} \gg \xi$, the trajectories of quasiparticles are dominated by random walks through the disordered landscape.  The diffusion constant of this random walk directly controls the momentum relaxation rate, and hence the resistivity.  One estimates that, up to a possibly small constant prefactor when the disorder potential is weak, the momentum relaxation time is proportional to the time it takes for a quasiparticle to travel across a puddle:
\begin{equation}
\rho \propto \frac{1}{\tau} \propto \frac{v_{\mathrm{F}}}{\xi}. \label{eq:mainbal}
\end{equation}
\red{Here we assume that the typical strength of the impurity potential is fixed and does not depend on $\xi$.}

\item \textbf{viscous hydrodynamic:}  In the limit where $\ell_{\mathrm{ee}} \ll \xi$, each quasiparticle collides with other quasiparticles in momentum-conserving collisions.  Suppose that when minimizing the entropy production in (\ref{eq:entcart}), we are able to arrange the quasiparticles in local thermodynamic equilibrium (on length scales small compared to $\xi$).  The time scale relevant for transport is the time it takes the collection of colliding quasiparticles to travel from the center of an impurity puddle to the edge -- this is how long it takes to feel the effects of inhomogeneity.    As the quasiparticles undergo random walks with diffusion constant $v_{\mathrm{F}}\ell_{\mathrm{ee}}$:
\begin{equation}
\rho \propto \frac{1}{\tau} \propto \frac{v_{\mathrm{F}} \ell_{\mathrm{ee}}}{\xi^2} \label{eq:mainvisc}
\end{equation}
In this regime, transport is governed by (a possibly generalized) viscous hydrodynamics \cite{KS06, KS11, Davison:2013txa, Lucas:2015lna, PhysRevB.93.075426}. \red{An understanding of how viscous hydrodynamics emerges from kinetic theory has been achieved in previous works \cite{levitov2,2016arXiv161200856L,levitov3}}.  Indeed, we will see in simple examples that the resistivity is proportional to the viscosity of the electron fluid. \red{In sections \ref{sec:genprin} and \ref{sec:bounds} in particular, we place these results in a broader formal structure. The resistivity (\ref{eq:mainvisc}) emerges whenever there are more inversion-odd conserved quantities than diffusive imbalance modes.}

\item \textbf{diffusive hydrodynamic:} \red{There is an obstruction to the emergence of the above viscous regime, even in the limit $\ell_{\mathrm{ee}} \ll \xi$.}  Suppose that there are many conserved quantities which are even under inversion symmetry (e.g. conserved scalar densities), and thus have odd currents;  but only a small number of odd conserved quantities in the absence of disorder (perhaps only momentum).    An example which we will study explicitly later in this paper is a theory of multiple Fermi surfaces, where the number of particles on each Fermi surface is conserved.  In general, the inhomogeneity makes it impossible to arrange the system to be in local thermodynamic equilibrium -- the number of constraints arising from the conservation laws associated with odd currents is too large to be solved with equilibrium values of the odd conserved quantities. Instead, the conservation laws may only be satisfied by creating ``non-equilibrium" imbalances of quasiparticles throughout the sample:  see Figure \ref{fig:imbfig}.  Locally, the fluid appears out of thermal equilibrium due to the presence of such currents.  Because quasiparticles collide after a distance $\ell_{\mathrm{ee}}$, the non-equilibrium quasiparticle gradient required to drive a current must be of order one quasiparticle per mean free path.  Hence, integrating this gradient over the impurity puddle, the number of quasiparticles excited out of equilibrium is $N_{\mathrm{imbalance}} \sim (\xi/\ell_{\mathrm{ee}})^2$.    We now estimate the resistivity by calculating the rate at which momentum is relaxed.  The electric field which drove one quasiparticle out of equilibrium in the viscous limit now drives $N_{\mathrm{imbalance}}$ quasiparticles out of equilibrium, and so 
\begin{equation}
\rho \propto \frac{1}{\tau_{\mathrm{viscous}}} \times N_{\mathrm{imbalance}}\propto \frac{v_{\mathrm{F}} \ell_{\mathrm{ee}}}{\xi^2} \times \frac{\xi^2}{\ell_{\mathrm{ee}}^2} \sim \frac{v_{\mathrm{F}}}{\ell_{\mathrm{ee}}}.  \label{eq:mainimb}
\end{equation}
The relaxational dynamics of the imbalance gradients is diffusive.
The existence of this novel transport regime is a main prediction of our theory. \red{Regimes of transport with a resistivity $\rho \propto 1/\ell_\text{ee}$, as in (\ref{eq:mainimb}), have previously arisen due to a thermal diffusive mode \cite{KS11} and also in compensated metals \cite{Baber383,levinson78}. The following two paragraphs elaborate on the relation of our result with these earlier works.}
\end{enumerate}

Let us discuss the diffusive hydrodynamic limit further. Firstly, in a conventional fluid, conservation of both energy and particle number leads to a thermal diffusion mode and hence a contribution to the resistivity similar to (\ref{eq:mainimb}).  However, the temperature dependence of the resistivity will not be governed entirely by the temperature dependence of $\ell_{\mathrm{ee}}$ in this case, but by factors of the entropy density that appear \cite{KS11}.  We emphasize the likely existence of non-thermal diffusion modes in many realistic metals which can lead to the resistivity (\ref{eq:mainimb}), with all temperature dependence in $\rho$ governed by the momentum-conserving scattering rate.  

Secondly, in cases where the extra conserved densities are due to the presence of multiple bands, one may ask whether or not (\ref{eq:mainimb}) is simply a re-derivation of Baber scattering \cite{Baber383}.  Baber scattering arises in a metal with two appreciably occupied bands, where one of the bands carries nearly all of the current, while the other band efficiently relaxes momentum. \red{Alternatively, electron and hole bands in the metal can precisely compensate each other so that the total charge density is zero, and hence charge dynamics is decoupled from momentum relaxation \cite{levinson78}.} A precise explanation of these effects can be found in \cite{maslov}. 
In these limits, one arrives at an equation similar to (\ref{eq:mainimb}).   In fact, the hydrodynamic mechanism for (\ref{eq:mainimb}) is significantly more general. \red{We require neither an asymmetry between the two bands nor that the total charge density vanish.}  

The resistivity (\ref{eq:mainimb}) readily admits a hydrodynamic interpretation.  It is possible to observe (\ref{eq:mainimb}) without long-lived quasiparticles, as we have emphasized in a companion paper \cite{ushydro}.   The advantage of the microscopic description of (\ref{eq:mainimb}) using kinetic theory is that we are able to further describe the crossover to a low temperature ballistic regime.   We are also able to describe the transition from ballistic to viscous hydrodynamic regimes, as well as models with multiple microscopic scattering rates, which exhibit all three regimes of transport, depending on the value of $\xi$.  Quantitatively characterizing crossovers between \red{the various classes of} hydrodynamic and ballistic transport is a \red{central} achievement of this work, \red{extending the recent results in \cite{levitov2,2016arXiv161200856L,levitov3} to a much broader class of models}.

Let us note that the list of transport phenomena we described is not fully exhaustive.   If there are not very many conservation laws to satisfy, one may be able to drive current along narrow contours in order to balance the effects of viscous dissipation and diffusive dissipation more efficiently than (\ref{eq:mainimb}) \cite{KS11}.  However, we expect that close to the ballistic-to-hydrodynamic crossover, which is where most realistic solid-state systems exist,  the most important effects  will be the three described above.

\subsection{Phenomenology for Experiments}
The hydrodynamic limit of transport has been directly observed in \cite{Molenkamp95, Bandurin1055, Crossno1058, Moll1061, 2017arXiv170306672K,WP2}.  Our proposal is that this limit may have \emph{already} been observed in a diverse group of materials, albeit in a subtle way -- through the unconventional temperature dependence of the resistivity.

Our main proposal is that the $T^2$ resistivity observed in (strongly) correlated Fermi liquids, and $T$-linear resistivity in the non-Fermi liquid regime of many such materials, including transition metal oxides \cite{husseyA, Hussey, Grigera329}, pnictides \cite{pnictide, Analytis2014},   heavy fermion metals \cite{Gegenwart2008,Si1161} and organic metals \cite{dressel}, has a common origin: a non-thermal diffusive mode limiting transport, as we have already described.   We now ask whether such a mode can exist in many strange metals.   The most natural imbalance mode is due to the imbalance of quasiparticles between different pockets and/or bands;  we will describe a toy model of this in detail in Section \ref{sec:2FS}.  Many strange metals (though not all) have complicated band structures, and this is a natural possibility.  For single band materials, such as cuprates \cite{Hussey} or $\mathrm{SrTiO}_3$ \cite{kamran, suzanne}, there are other possible imbalance modes,  including spin imbalance modes\footnote{A careful treatment of spin imbalance diffusion is beyond this work for technical reasons (time-reversal symmetry is broken), but may exhibit similar behavior to a model of two Fermi surfaces.}, `quadrupole' fluctuations (section \ref{sec:IN}) and modes associated to additional degrees of freedom such as phonons.    

Our proposal requires that the impurity potential be smooth.   This is actually rather natural to obtain in many strange metals, which are quasi-two-dimensional layered materials, with clean conduction layers separated by a distance $d\sim 1.5$ nm from dirty dopant layers.  The static Coulomb potential created by the random arrangement of ions will be random and vary significantly only on length scales larger than \begin{equation}
\xi \sim d\times \sqrt{\frac{\epsilon_\parallel}{\epsilon_\perp}},
\end{equation}
where  $\epsilon_\parallel$ is an `effective dielectric constant' associated to in-plane electric fields, and $\epsilon_\perp$ is the `effective dielectric constant' for out-of-plane electric fields.  This equation straightforwardly follows from Gauss' law in an anisotropic medium.   In the monolayer cuprates, we estimate $\epsilon_\parallel/\epsilon_\perp \approx 1.2$ \cite{1998RvMP...70..897K},  though we caution that this is simply an order of magnitude estimate;   other materials may be more anisotropic.    Strange metals can have mean free paths as short as $\ell_{\mathrm{ee}} \sim 1 $ nm \cite{RevModPhys.75.1085, MIR}, which could be smaller than $\xi$.  Even slightly outside the quantum critical fan of such bad metals one can expect $\ell_{\mathrm{ee}} \sim \xi$, and so our hydrodynamic mechanism continues to describe transport, both in the strange metal and in the Fermi liquid.   Furthermore, in many strange metals, the amplitude of the disorder, which is related to the dopant concentration, is not tunable without moving out of the non-Fermi liquid regime.  One cannot arbitrarily reduce the disorder.   Thus, we expect that the disorder is \emph{large amplitude and long wavelength} in realistic materials.  This is precisely the regime where diffusion-limited transport naturally occurs.

Finally, we note that many materials, including very pure atomic metals like Au or Pb \cite{1934Phy1.1115D} and doped $\mathrm{SrTiO}_3$ \cite{suzanne} exhibit sharp downturns of a few percent in the resistivity at low temperature.   These downturns cannot be associated with the Kondo effect because $\partial \rho / \partial T$ is not vanishing as $T\rightarrow 0$.   This downturn is consistent with viscous effects.  We will describe in Section \ref{sec:2FS} a microscopic toy model with phenomenology very similar to these materials.   Furthermore, our formalism elucidates why many metals which are believed to be clean and strongly correlated do not exhibit obvious signatures of viscous transport in bulk resistivity measurements -- there may be additional non-thermal diffusive modes whose contributions to the resistivity overwhelm viscous effects.  It would be interesting to revisit these phenomena in more detail, using our \red{more} complete kinetic theory of transport.

\section{The Boltzmann Equation}
\label{sec:boltz}
We now turn to the detailed solution of the kinetic theory of transport.  Let us consider weakly interacting fermionic quasiparticles with an effective dispersion relation $\epsilon(\vec p)$, in the absence of disorder.  We assume that in a perfectly clean sample, in thermal equilibrium, the distribution function is  given by the non-interacting Fermi function:  \begin{equation}
f_{\mathrm{eq}} = n_{\textsc{f}}\left(\frac{\epsilon(\vec p) - \mu_0}{T}\right), \;\;\;\; n_{\textsc{f}}(x) = \frac{1}{1+\mathrm{e}^{x}}.   \label{eq:fermi}
\end{equation}
We will only use the specific form of $f_{\mathrm{eq}}$ to estimate the temperature dependence of certain coefficients (for conventional Fermi liquids).    We will neglect the effects of the underlying lattice, other than through their modification of the band structure $\epsilon(\vec p)$ -- hence, we will not include phonons in our kinetic theory, nor will we account for umklapp.  


It has long been known that in such a Fermi liquid, so long as the charge density is finite,  the resistivity vanishes.   In order to obtain a nonzero resistivity, we will suppose that the chemical potential is inhomogeneous:   \begin{equation}
\mu(\vec x) = \mu_0 - V_{\mathrm{imp}}(\vec x).
\end{equation}
The equilibrium distribution function is now $\vec x$-dependent. When the inhomogeneity length scale $\xi$ is long compared to the Fermi wavelength $\lambda_{\mathrm{F}}$ then \begin{equation}
f_{\mathrm{eq}}(\vec x, \vec p) = n_{\textsc{f}}\left(\frac{\epsilon(\vec p) - \mu(\vec x)}{T}\right).
\end{equation}
Here and henceforth we have set $k_{\mathrm{B}}=1$.

Now, we apply an infinitesimal electric field $\vec E$, which will perturb the true distribution function $f$ a bit away from $f_{\mathrm{eq}}$.  We will then solve the Boltzmann equation in order to compute the resistivity:
\begin{equation}
\frac{\partial f}{\partial t} + \vec v \cdot \frac{\partial f}{\partial \vec x} + (\vec F -e \vec E)\cdot \frac{\partial f}{\partial \vec p} = -\mathcal{C}[f]\,.  \label{eq:boltzmann}
\end{equation}
In this equation, $\vec v$ is the quasiparticle velocity \begin{equation}
\vec v \equiv \frac{\partial \epsilon}{\partial \vec p}\,,
\end{equation}
and $\vec F$ is the external force from the impurity potential: \begin{equation}
\vec F \equiv -\frac{\partial V_{\mathrm{imp}}}{\partial \vec x}\,.
\end{equation}
$\mathcal{C}$ is a local collision term subject to suitable conservation laws, and respecting Fermi-Dirac statistics of a weakly-interacting quantum gas of fermions.  In particular, $f_{\mathrm{eq}}$ must be an exact solution of (\ref{eq:boltzmann}) when $\vec E=0$.   This implies that the collision operator $\mathcal{C}$ has zeros associated with the local conservation laws of charge and energy.    More complicated disorder which couples to $f$ with more than a simple $p$-derivative corresponds to impurity potentials that couple to other operators in the QFT.  We remind the reader that the collision operator can be thought of as encoding the decay rate of the quasiparticles: crudely speaking,\begin{equation}
\mathcal{C}[f] \sim \mathrm{Im}\left(\Sigma[f]\right) f.
\end{equation}
with more precise expressions found in \cite{kamenev}.

Let us briefly remind the reader of the assumptions going into (\ref{eq:boltzmann}) \cite{kamenev}.   The Boltzmann equation can be rigorously derived from the Schwinger-Keldysh formalism, and is a controlled expansion when (\emph{i}) the scales over which $f(x,p)$ varies obey $|\mathrm{\Delta}x| \cdot |\mathrm{\Delta}p| \gg \hbar$.   For our purposes this will correspond to $|\mathrm{\Delta}x| \gg \lambda_{\mathrm{F}}$ -- hence, we will assume that the function $V_{\mathrm{imp}}(\vec x)$ is smooth on microscopic scales;  (\emph{ii}) quasiparticles are well-defined, which qualitatively means that all scattering rates (the eigenvalues of the linearized $\mathcal{C}$ operator) are all small compared to $k_{\mathrm{B}}T/\hbar$.   In such a limit, the collision operator will likely be well-approximated by a small number of Feynman diagrams and can be computed,  although we will not do so explicitly at any point in this paper.    We will also neglect renormalization of $\epsilon$ and $V_{\mathrm{imp}}$ over their bare values, due to quantum fluctuations,  though this can be accounted for \cite{kamenev}.    

Our goal is to find stationary solutions to the kinetic equations to linear order in $\vec E$.   We write $f\approx f_{\mathrm{eq}} + \mdelta f + \mathcal{O}(E^2)$.  Because $f_{\mathrm{eq}}$ is an exact solution to the kinetic equations, up to the electric field  contributions, we obtain at leading order:  \begin{equation}
\vec v \cdot \partial_x \mdelta f + \vec F \cdot \partial_p \mdelta f +e \vec E \cdot \vec v \left(-\frac{\partial f_{\mathrm{eq}}}{\partial \epsilon}\right) = \left.\frac{\mdelta \mathcal{C}}{\mdelta f}\right|_{f=f_{\mathrm{eq}}}\mdelta f.
\end{equation}
This equation is a classical linear differential equation, and we will heavily employ the technology of linear algebra.   In order to do so most efficiently, it is helpful to write this equation in terms of a variable $\Phi$, defined via \begin{equation}
\mdelta f \equiv \left(-\frac{\partial f_{\mathrm{eq}}}{\partial \epsilon}\right)\Phi.
\end{equation}
We then interpret $\Phi(\vec x, \vec p)$ as a vector $|\Phi\rangle$ in an infinite dimensional vector space:  \begin{equation}
|\Phi\rangle \equiv \int \mathrm{d}^dx \mathrm{d}^dp \; \Phi(\vec x, \vec p) |\vec x \vec p\rangle.
\end{equation}
Let us define an inner product \begin{equation}
\langle \vec x \vec p |\vec x_0 \vec p_0\rangle \equiv \frac{1}{(2\pi\hbar)^dV_x}\left(-\frac{\partial f_{\mathrm{eq}}(\vec x, \vec p)}{\partial \epsilon}\right) \mdelta(\vec x - \vec x_0) \mdelta(\vec p - \vec p_0) \,, \label{eq:innerproduct}
\end{equation}
with $V_x$ the spatial volume of the theory.  While the distribution function $f(\vec x, \vec p)$ is real, we will sometimes Fourier transform the spatial coordinate $\vec x$, and in this case the inner product above should be understood as complex.  (\ref{eq:innerproduct}) is useful because with a sharp Fermi surface,  the distribution function $\mdelta f$ is generally quite singular and sharply peaked around the Fermi surface.   The functions $\Phi$ are smooth functions, in contrast.   Furthermore, the weighted inner product (\ref{eq:innerproduct}) will not diverge on any sensible (polynomial in $\vec p$) trial function.

With the definitions above we can write the
linearized Boltzmann equation in the abstract form
\begin{equation}
(\mathsf{W} + \mathsf{L}) |\Phi\rangle  = E_i |\mathsf{J}_i\rangle.  \label{eq:LAmain}
\end{equation}
Here we introduced the streaming operator \begin{equation}
\mathsf{L}|\vec x \vec p\rangle \equiv- \int \mathrm{d}^dx_0 \mathrm{d}^dp_0  \left(\vec v\cdot \partial_x + \vec F \cdot \partial_p\right)\mdelta(\vec x-\vec x_0)\mdelta(\vec p - \vec p_0)  \left|\vec x_0 \vec p_0 \right\rangle, \label{eq:Ldef}
\end{equation}
the linearized collision operator \begin{equation}
\mathsf{W}|\vec x \vec p\rangle \equiv \int \mathrm{d}^dp_0 \frac{\mdelta \mathcal{C}(x,p)}{\mdelta f(x,p_0)} |\vec x \vec p_0\rangle ,
\end{equation}
and the source vector \begin{equation}
|\mathsf{J}_i\rangle \equiv -e \int \mathrm{d}^dx \mathrm{d}^dp \;  v_i(\vec p)|\vec x \vec p\rangle.  \label{eq:Edef}
\end{equation}

$\mathsf{L}$ is an antisymmetric matrix:  when integrating by parts across $\partial_\epsilon f_{\mathrm{eq}}$ in (\ref{eq:innerproduct}), the $x$ and $p$ derivatives can easily be shown to cancel.  We will further assume that the microscopic kinetic theory is time reversal symmetric and inversion symmetric in this paper;  the latter assumption requires that the band structure obey $\epsilon(\vec p) = \epsilon(-\vec p)$.  Under these assumptions, $\mathsf{W}$ is a symmetric matrix in $\vec p$ \cite{ziman}:  time reversal invariance implies $\mathsf{W}(\vec p, \vec q) = \mathsf{W}(-\vec q, -\vec p)$ while inversion symmetry implies $\mathsf{W}(\vec p, \vec q) = \mathsf{W}(-\vec p, -\vec q)$.  For simplicity, we have suppressed $\vec x$ indices in these equations, and for the remainder of the paragraph.  The assumption that $f_{\mathrm{eq}}$ describes a stable thermal equilibrium implies that $\mathsf{W}$ is positive semidefinite.   Upon decomposing $|\Phi\rangle$ into its even/odd components in $\vec p$, the matrices $\mathsf{W}$ and $\mathsf{L}$ are restricted to the following sectors:  \begin{equation}
|\Phi\rangle = \left(\begin{array}{c} |\Phi_{\mathrm{e}}\rangle \\ |\Phi_{\mathrm{o}}\rangle \end{array}\right), \;\;\;\;\; \mathsf{W} = \left(\begin{array}{cc}  \mathsf{W}_{\mathrm{ee}} &\ 0 \\ 0 &\ \mathsf{W}_{\mathrm{oo}} \end{array}\right), \;\;\;\;\; \mathsf{L} = \left(\begin{array}{cc}  0 &\ \mathsf{L}_{\mathrm{eo}}  \\  \mathsf{L}_{\mathrm{oe}} &\ 0 \end{array}\right).   \label{eq:WLEO}
\end{equation}
$\mathsf{W}_{\mathrm{ee}}$ and $\mathsf{W}_{\mathrm{oo}}$ are symmetric, and $\mathsf{L}_{\mathrm{eo}} = -\mathsf{L}_{\mathrm{oe}}^{\mathsf{T}}$.  The fact that $\mathsf{W}$ is block diagonal follows from the fact that the odd-even block vanishes (due to time reversal symmetry).  For an even perturbation $|\Phi\rangle$,  $\mathsf{W}|\Phi\rangle$ has no odd component: \begin{align}
\int \mathrm{d}^dq \; [\mathsf{W}(\vec p, \vec q) - \mathsf{W}(-\vec p, \vec q)] \Phi(\vec q) &= \int \mathrm{d}^dq \; [\mathsf{W}(\vec p, \vec q) - \mathsf{W}(\vec p, -\vec q)] f(\vec q) \notag \\
&= \int \mathrm{d}^dq \; [\mathsf{W}(\vec p, \vec q)f(\vec q) - \mathsf{W}(\vec p, -\vec q)f(-\vec q)] = 0.
\end{align}  
  The last step follows from the fact that $\vec q$ is a dummy integration variable.

The matrix $\mathsf{W}$ will also have null vectors associated with conservation laws.   To see this, we note that if the quantity $\int \mathrm{d}^d\vec p \mathrm{d}^d\vec x  \, \alpha(\vec p)f$ is conserved, then there must be a \emph{family} of solutions to the nonlinear Boltzmann equation associated with shifts in the conjugate thermodynamic variable $\zeta$:
\begin{equation}
f_{\mathrm{eq}}\left(\frac{\epsilon}{T}\right) \rightarrow f_{\mathrm{eq}}\left(\frac{\epsilon - \zeta \cdot \alpha(\vec p)}{T}\right).
\end{equation}
For example, setting $\alpha(\vec p) = 1$, we obtain a conservation law for electron number;  the conjugate thermodynamic variable is the chemical potential $\mu$.   
The linearized collision operator must have a zero mode associated with the fact that $\partial_\zeta \mathcal{C}[f_{\mathrm{eq}}(\zeta)] = 0$: \begin{equation}
\mathsf{W} \int \mathrm{d}^dp \; \alpha(\vec p) | \vec x \vec p\rangle = 0.
\end{equation}

Because the charge current is locally given by \begin{equation}
\vec J(\vec x)  =  -e \int \mathrm{d}^d p  \vec v(\vec p) f,
\end{equation}
we find \begin{equation}
\langle \Phi |\mathsf{J}_i\rangle = -\frac{e}{V_x} \int \frac{\mathrm{d}^dx \mathrm{d}^dp}{(2\pi\hbar)^d}    v_i(\vec p)  \left(-\frac{\partial f_{\mathrm{eq}}}{\partial \epsilon} \right) \Phi  = \frac{1}{V_x} \int \mathrm{d}^dx  J_i.  \label{eq:getcurrent}
\end{equation}
Hence, the source vector $|\mathsf{J}_i\rangle$ is the `basis vector' for the homogeneous part of the electric current.

Our goal is to compute the resistivity tensor, which is given by $E_i = \rho_{ij}J_j$.  Alternatively, we may compute the conductivity tensor $J_i = \sigma_{ij}E_j$.      Using (\ref{eq:getcurrent}), we see that  \begin{equation}
\sigma_{ij} = \langle\mathsf{J}_i|(\mathsf{W}+\mathsf{L})^{-1}|\mathsf{J}_j\rangle . \label{eq:JrhoJ}
\end{equation}
Unfortunately, in general, one cannot perform the inverse $(\mathsf{W}+\mathsf{L})^{-1}$ explicitly.   In Section \ref{sec:perturb}, we will show how to invert $\mathsf{W}+\mathsf{L}$ analytically, at leading perturbative order in $V_{\mathrm{imp}}$.  In Section \ref{sec:bounds}, we will develop a variational technique that allows for non-perturbative upper bounds on the resistivity.

Historically, one neglects the streaming terms and approximates $\mathsf{L} = 0$ \cite{ziman}.  However, we clearly cannot do this in our model -- we have neglected umklapp scattering and hence momentum is conserved in electron-electron collisions. Therefore the vector \begin{equation}
|\mathsf{P}_i, \vec x\rangle \equiv \int \mathrm{d}^dp \; p_i |\vec x \vec p\rangle \label{eq:PhiP}
\end{equation} is a null vector of $\mathsf{W}$:  $\mathsf{W}|\mathsf{P}_i, \vec x \rangle = 0$ for all $\vec x$.  This is sufficient to prove that the conductivity is infinite, and hence $\rho=0$, if the streaming terms can be neglected.

Another approximation which has been made is to neglect interactions ($\mathsf{W}=0$) and treat inhomogeneity perturbatively \cite{mirlin}.  We will review this approach in Section \ref{sec:free}, but our more general formalism allows for a complete treatment of both inhomogeneity and interactions.

\section{Perturbation Theory for Weak Disorder}
\label{sec:perturb}

\subsection{General Considerations}
So far, our comments have been quite general.  However, it is useful to have precise quantitative results for $\rho$ across the ballistic-to-hydrodynamic crossover.  Towards this end,  we will completely solve the transport problem at weak disorder.  More precisely, suppose that (upon disorder averaging, denoted with $\mathbb{E}$): \begin{equation}
\mathbb{E}[V_{\mathrm{imp}}] = 0, \;\;\; \mathbb{E}\left[V_{\mathrm{imp}}(\vec x)V_{\mathrm{imp}}(\vec y) \right] = \delta^2 \mathcal{F}\left(\frac{|\vec x - \vec y|}{\xi}\right).  \label{eq:Vimp}
\end{equation}
Here $\mathcal{F}$ is an $\mathcal{O}(1)$ function, and $\xi$ is a length scale associated with the disorder distribution.  We will define the Fourier transform \begin{equation}
V_{\mathrm{imp}}(\vec k) \equiv \frac{1}{\sqrt{V_x}} \int \mathrm{d}^dx \; \mathrm{e}^{\mathrm{i}\vec k \cdot \vec x} V_{\mathrm{imp}}(\vec x),
\end{equation}
and typically assume that for $k\rightarrow \infty$,  $V_{\mathrm{imp}}(k) \sim \exp[-k\xi]$.   The volume-dependent prefactor appearing above is such that $\int \mathrm{d}^dk |V_{\mathrm{imp}}(\vec k)|^2 = \delta^2$, and will simplify notation shortly.  If $\delta$ is a perturbatively small parameter, then at leading order (so long as the charge density is not zero in equilibrium) one finds the resistivity $\rho = \rho_2 \delta^2 + \mathrm{O}(\delta^4)$.   We will exactly compute the coefficient $\rho_2$, making the simple assumption that the only locally conserved quantity, odd under inversion, is momentum.  We relax this assumption in (\ref{eq:multcons}).

As this section is a little technical, we state the result upfront. We will find that to leading nontrivial order in the impurity potential, the resistivity
\begin{equation}\label{eq:rhoij}
\rho_{ij} = \frac{1}{n^2e^2} \int \frac{\mathrm{d}^dk}{(2\pi)^d} k_i k_j |V_{\mathrm{imp}}(k)|^2 \mathcal{A}(k), 
\end{equation}
where we define \begin{equation}
\mathcal{A}(k)\equiv \langle \mathsf{n}| (\mathsf{W}+\mathsf{L})^{-1}_{\vec k, \mathrm{clean}}  | \mathsf{n}\rangle.  \label{eq:sigmaexact4}
\end{equation}
We have denoted $|\mathsf{n}, \vec k = \vec 0 \rangle = |\mathsf{n}\rangle$, where $|\mathsf{n}, \vec k \rangle$ will be defined in (\ref{eq:1k}) below. 

In addition to the derivation that follows,  in Appendix \ref{app:MM} we re-derive the expression (\ref{eq:sigmaexact4}) from a  more sophisticated (quantum-mechanical) framework called the memory matrix formalism.  Using this formalism, we learn that $\mathcal{A}(k)$ is the spectral weight of the density operator, evaluated (in our case) in a classical kinetic limit:  it tells us how efficiently we may lose momentum off of impurities on the length scale $k^{-1}$.

We may just as well compute the conductivity $\sigma \sim \delta^{-2}$.  In order to do so, we employ the following key observation.  When the disorder is perturbatively small, it is instructive to Fourier transform the position coordinate $\vec x$ to a wave number coordinate $\vec k$.  We write $\mathsf{W}$ and $\mathsf{L}$ in the following block-diagonal form (this is different and unrelated to the block diagonal form mentioned in (\ref{eq:WLEO})): \begin{equation}
\mathsf{W}+ \mathsf{L} = \left(\begin{array}{cc}  (\mathsf{W}+\mathsf{L})_{\vec 0, \vec 0} &\ (\mathsf{W}+\mathsf{L})_{\vec 0, \vec k} \\ (\mathsf{W}+\mathsf{L})_{\vec k^\prime, \vec 0}  &\ (\mathsf{W}+\mathsf{L})_{\vec k^\prime, \vec k} \end{array}\right) \propto \left(\begin{array}{cc}  1 &\ \delta \\ \delta  &\ 1 \end{array}\right).  \label{eq:WL00block}
\end{equation}
Only the diagonal pieces are non-vanishing at $\delta=0$, when momentum is conserved.  The exact conductivity is given by (\ref{eq:JrhoJ}).  By construction in  (\ref{eq:Edef}), $|E\rangle$ is only non-vanishing in the $\vec k = \vec 0$ sector.   By (\ref{eq:JrhoJ}), we are looking for an eigenvalue of $(\mathsf{W}+\mathsf{L})^{-1}$, overlapping with $|E\rangle$, which diverges as $\delta^{-2}$.   So we break up the spatially homogeneous $\vec0\vec0$ block of (\ref{eq:WL00block}) into a further $2\times 2$ block by separating out the null eigenvectors of $\mathsf{W}$:
\begin{equation}
(\mathsf{W}+\mathsf{L})_{\vec 0, \vec 0} = \left(\begin{array}{cc} 0 &\ 0 \\ 0 &\ \mathsf{W}_0 + \mathrm{O}\left(\delta^2\right) \end{array}\right).  \label{eq:W0d2}
\end{equation}
To obtain this form, let us consider for simplicity the momentum (abstract) vectors \begin{equation}
|\mathsf{P}_i, \vec k\rangle = \int \mathrm{d}^dx \mathrm{d}^dp  \; p_i \mathrm{e}^{\mathrm{i}\vec k \cdot \vec x}|\vec x \vec p\rangle.
\end{equation}
From (\ref{eq:PhiP}), $\mathsf{W} | \mathsf{P}_i, \vec k = \vec 0\rangle = 0$, to all orders in $\delta$.  Similarly, \begin{equation}
\mathsf{L}|\mathsf{P}_i, \vec k\rangle = \int \mathrm{d}^dx \mathrm{d}^dp \;  \left((\mathrm{i}\vec v(\vec p) \cdot \vec k)p_i  - \frac{\partial V_{\mathrm{imp}}}{\partial x_i} \right) \mathrm{e}^{\mathrm{i}\vec k \cdot \vec x} |\vec x \vec p\rangle.  \label{eq:eqLonly}
\end{equation}
When $\vec k =\vec 0$,  the first term vanishes, and the second term has no homogeneous component.  The inner product with a $\vec k =\vec 0$ vector would be proportional to   $\int \mathrm{d}^dx \; (-\partial_\epsilon f_{\mathrm{eq}})\nabla V_{\mathrm{imp}} = \int \mathrm{d}^dx \; \nabla f_{\mathrm{eq}} = 0$.   This explains why (\ref{eq:W0d2}) holds to all orders in $\delta$.

Using block diagonal matrix inversion identities, and for simplicity denoting \begin{equation}\label{eq:Adef}
(\mathsf{W}+\mathsf{L})_{\vec 0, \vec k} = \delta \left(\begin{array}{c} \mathsf{A}_1 \\ \mathsf{A}_2 \end{array}\right), \;\;\;\;\; (\mathsf{W}+\mathsf{L})_{\vec k^\prime, \vec 0} = \delta \left(\begin{array}{cc} \mathsf{A}_3 &\ \mathsf{A}_4 \end{array}\right),
\end{equation}we obtain  
\begin{equation}
\left[(\mathsf{W}+\mathsf{L})^{-1}\right]_{\vec0\vec0} \approx \left[\left(\begin{array}{cc} 0 &\ 0 \\ 0 &\ \mathsf{W}_0 \end{array}\right) - \delta^2\left(\begin{array}{c} \mathsf{A}_1 \\ \mathsf{A}_2 \end{array}\right) (\mathsf{W}+\mathsf{L})_{\vec k^\prime, \vec k}^{-1} \left(\begin{array}{cc} \mathsf{A}_3 &\ \mathsf{A}_4 \end{array}\right)  \right]^{-1} \propto \left(\begin{array}{cc} \delta^2 &\ \delta^2 \\ \delta^2 &\ 1 \end{array}\right)^{-1}.
\end{equation}
Using the $\delta$-scalings in this equation, we can easily see that to leading order as $\delta \rightarrow 0$, only one sub-block of $\left[(\mathsf{W}+\mathsf{L})^{-1}\right]_{\vec0\vec0}$ is divergent:
\begin{equation}
\left[(\mathsf{W}+\mathsf{L})^{-1}\right]_{\vec0\vec0} = \left(\begin{array}{cc} -\delta^{-2}(\mathsf{A}_1 (\mathsf{W}+\mathsf{L})_{\vec k^\prime, \vec k}^{-1} \mathsf{A}_3 )^{-1} &\ \mathrm{O}(\delta^0) \\ \mathrm{O}(\delta^0) &\ \mathrm{O}(\delta^0) \end{array}\right)  + \cdots .
\end{equation}
Hence
\begin{equation}
\sigma_{ij} \approx \langle \mathsf{J}_i | \left(\begin{array}{cc} -\delta^{-2}(\mathsf{A}_1 (\mathsf{W}+\mathsf{L})_{\vec k^\prime, \vec k}^{-1} \mathsf{A}_3 )^{-1} &\ \mathrm{O}(\delta^0) \\ \mathrm{O}(\delta^0) &\ \mathrm{O}(\delta^0) \end{array}\right) |\mathsf{J}_j\rangle.  \label{eq:41sigma}
\end{equation}

We now must compute $\mathsf{A}_{1,3}$ to leading order in $\delta$.  The first observation that we make is that we may neglect the $V_{\mathrm{imp}}$ dependence in the   inner product (\ref{eq:innerproduct});  we have already extracted the leading order $\delta$-dependence and hence the  $V_{\mathrm{imp}}$-dependence of the inner product will only contribute to the conductivity at subleading orders.  Furthermore, $\mathsf{A}_{1,3}$ must come entirely from the streaming terms because, by definition in (\ref{eq:Adef}), they correspond to the null space of $\mathsf{W}$.  Hence, $\mathsf{A}_1 = -\mathsf{A}_3^{\mathsf{T}}$.  Because we have assumed that momentum is the only odd conserved quantity,  the only inversion-odd vectors $\mathsf{A}_{1,3}$ project on to are $|\mathsf{P}_i, \vec x\rangle$ (or $|\mathsf{P}_i, \vec k\rangle$).   We conclude that to leading order in $\delta$: \begin{equation}
\sigma_{ij} \approx \langle \mathsf{J}_i | \mathsf{P}_k\rangle \frac{1}{\langle \mathsf{P}_k| \mathsf{L}_{\vec 0 \vec k} (\mathsf{W} + \mathsf{L})^{-1}_{\vec k \vec k^\prime, \text{clean}} \mathsf{L}_{\vec k^\prime \vec 0} |\mathsf{P}_l\rangle} \langle \mathsf{P}_l | \mathsf{J}_j\rangle.   \label{eq:sigmaexact40}
\end{equation}
We have denoted $|\mathsf{P}_i, \vec k = \vec 0\rangle = |\mathsf{P}_i\rangle$ for simplicity.
The notation $(\mathsf{W} + \mathsf{L})^{-1}_{\vec k \vec k^\prime, \text{clean}}$ reminds us that, because we are only computing to leading order in $\delta$, we may approximate $(\mathsf{W}+\mathsf{L})^{-1}$ with its value when $\delta=0$.   


Let us now simplify (\ref{eq:sigmaexact40}).  Firstly, \begin{align}
    \langle\mathsf{J}_i | \mathsf{P}_j\rangle &= -e\int \frac{\mathrm{d}^dp}{(2\pi\hbar)^d} \, p_i v_j \left(-\frac{\partial f_{\mathrm{eq}}}{\partial \epsilon}\right) = e\int \frac{\mathrm{d}^dp}{(2\pi\hbar)^d}  \, p_i \frac{\partial \epsilon}{\partial p_j} \frac{\partial f_{\mathrm{eq}}}{\partial \epsilon}  = -e\delta_{ij} \int \frac{\mathrm{d}^dp}{(2\pi\hbar)^d} f = -\delta_{ij}ne.  \label{eq:EPn}
\end{align}
Secondly, \begin{equation}
\mathsf{L}_{\vec k \vec 0} |\mathsf{P}_i,\vec k = \vec 0\rangle = F_i|\mathsf{n},\vec k\rangle = \mathrm{i}k_i V_{\mathrm{imp}}(\vec k) |\mathsf{n},\vec k\rangle,
\end{equation}
where \begin{equation}\label{eq:1k}
|\mathsf{n},\vec k\rangle \equiv \int \mathrm{d}^dx \mathrm{d}^dp \; \mathrm{e}^{\mathrm{i}\vec k \cdot \vec x}|\vec x \vec p\rangle.
\end{equation}    
Combining (\ref{eq:sigmaexact40}) and (\ref{eq:EPn}), we find the resistivity (\ref{eq:rhoij}).

The remainder of this section provides a detailed analysis of (\ref{eq:sigmaexact4}) in various solvable limits of kinetic theory.   In this perturbative limit, we will be able to completely and unambiguously characterize the consequences of interactions on transport for the first time.

\subsection{Non-Interacting Theory}
\label{sec:free}

We begin by analyzing a non-interacting theory where $\mathsf{W}=0$ -- the only dynamics comes from the streaming terms.   Actually, it is important to keep $\mathsf{W}$ as an infinitesimal regulator $\mathsf{W} \sim z>0$.  This is because in the clean theory, $\mathsf{L} =  \mathrm{i}\vec k \cdot \vec v(\vec p)$,
and hence 

\begin{align}
\rho_{ij} &= \frac{1}{n^2e^2} \int \frac{\mathrm{d}^dp}{(2\pi\hbar)^d} \left[-\frac{\partial f_{\mathrm{eq}}}{\partial \epsilon}\int \frac{\mathrm{d}^dk}{(2\pi)^d} k_i k_j |V_{\mathrm{imp}}(k)|^2 \frac{1}{z+\mathrm{i}\vec k \cdot \vec v(\vec p)} \right] \notag \\
&= \frac{1}{n^2e^2} \int \frac{\mathrm{d}^dp}{(2\pi\hbar)^d} \left[-\frac{\partial f_{\mathrm{eq}}}{\partial \epsilon}\int \frac{\mathrm{d}^dk}{(2\pi)^d} k_i k_j |V_{\mathrm{imp}}(k)|^2 \mathrm{Re}\left(\frac{1}{z+\mathrm{i}\vec k \cdot \vec v(\vec p)}\right) \right] \notag \\
&= \frac{\pi}{n^2e^2} \int \frac{\mathrm{d}^dp}{(2\pi\hbar)^d} \left[-\frac{\partial f_{\mathrm{eq}}}{\partial \epsilon}\int \frac{\mathrm{d}^dk}{(2\pi)^d} k_i k_j |V_{\mathrm{imp}}(k)|^2 \delta(\vec k \cdot \vec v(\vec p))  \right] 
\label{eq:42res}
\end{align}
In the second step, we have used that $|V_{\mathrm{imp}}(k)|^2$ is an even function of $\vec k$.  In the third step we have taken the regulator $z\rightarrow 0$.   The factor of $-\partial f_{\mathrm{eq}}/\partial \epsilon$ comes from the inner product (\ref{eq:innerproduct}). 

Let us begin by evaluating this in the limit $T\rightarrow 0$.  In this limit \begin{equation}
-\frac{\partial f_{\mathrm{eq}}}{\partial \epsilon} = \delta(\epsilon-\mu).
\end{equation}
For simplicity in what follows, let us also assume that the Fermi surface is spherically symmetric.  While this is not generally true, relaxing this assumption leads to angular prefactors alone.  Upon performing the angular integral over $p$ we obtain \begin{equation}
\rho_{ij} \propto \frac{1}{n^2e^2} \int \frac{\mathrm{d}^{d}k}{(2\pi)^{d-1}} k_i k_j |V_{\mathrm{imp}}(k)|^2 \frac{\nu(\mu)}{|k| v_{\mathrm{F}}},  \label{eq:42result}
\end{equation}
where we have neglected an overall constant prefactor.  $\nu(\mu)$ is the density of states at the Fermi surface: \begin{equation}
    \nu(\mu) = \int \frac{\mathrm{d}^dp}{(2\pi\hbar)^d} \delta(\epsilon(\vec p)-\mu).
\end{equation} 
Employing (\ref{eq:Vimp}), and noting rotational invariance of the disorder, we obtain \begin{equation}
\rho_{ij}  \propto \frac{\nu(\mu) \delta^2 }{n^2e^2 v_{\mathrm{F}} \xi}   \delta_{ij}.
\end{equation}
For a spherical Fermi surface, we have $n\propto p_{\mathrm{F}}^d$ and $\nu(\mu)v_{\mathrm{F}} \propto p_{\mathrm{F}}^{d-1}$.  Defining $m\equiv p_{\mathrm{F}}/v_{\mathrm{F}}$,  $\delta = m v_{\mathrm{F}}^2 \theta$,  where $\theta$ is roughly the angle a quasiparticle is scattered by the disorder on length scales of order $\xi$, we obtain  \begin{equation}
\rho \propto  \frac{m}{ne^2\tau_{\mathrm{imp}}}
\end{equation}
where $\tau_{\mathrm{imp}} \equiv \xi/v_{\mathrm{F}}\theta^2$ is the momentum relaxation time.    This is nothing more than the canonical formula for the residual resistivity \cite{mirlin2}, in the limit of small-angle scattering.   We emphasize that the scaling\footnote{We have assumed -- \red{as noted under equation (\ref{eq:mainbal}) above} -- that the typical strength of the impurity potential $\delta$ has been kept fixed.  In most previous literature, $\delta$ is defined in a $\xi$-dependent way.} \begin{equation}
\rho \propto \frac{1}{\xi}
\end{equation}
is a universal consequence of this \red{ballistic} limit.   One of our \red{main concerns} will be the breakdown of this scaling due to electron-electron interactions.

Not surprisingly, we have found a residual resistivity due to impurity scattering.  On closer inspection, this is slightly subtle:  we previously formulated a bound on the resistivity associated with entropy production.  The key point is that the regulator $z$ that we imposed in (\ref{eq:42res}) is sufficient to lead to  ``spontaneous" production of entropy:  $(z+\mathsf{L})^{-1}$ has a non-vanishing symmetric component.   We associate this entropy production with the emergence of an  ``arrow of time".   Alternatively, we note that the microscopic trajectories of single particles in random potentials in spatial dimensions $d\ge 2$ are diffusive:  in certain limits, this has been proven rigorously \cite{papanicolaou, lebowitz}.   The computation that we have done is a  perturbative computation of the associated diffusion constant, which we can obtain from $\rho_{ij}$ via an Einstein relation.  

\subsubsection{Thermal Effects}
Before moving on to account for electron-electron interactions, let us briefly mention thermal corrections to this residual resistivity.  At very low temperatures, we employ a Sommerfeld expansion of $(-\partial f_{\mathrm{eq}}/\partial \epsilon)$ in (\ref{eq:42res}).   If, for simplicity, we retain the assumption of spherical symmetry and assume that $n$ is held fixed with increasing $T$, then, using that \begin{equation}
    n(\mu,T) \approx n_{T=0}(\mu) + \frac{\pi^2T^2}{6}\nu^\prime(\mu) + \cdots 
\end{equation}we find \begin{equation}
\mu(T) \approx \mu_0 - \frac{\nu^\prime(\mu)}{\nu(\mu)}\frac{\pi^2T^2}{6}.
\end{equation}
If we neglect the $k$-dependence in $|V_{\mathrm{imp}}(k)|^2$, we find, using the Sommerfeld expansion: \begin{align}
\rho(T) &\propto \frac{1}{n^2e^2} \int \frac{\mathrm{d}^dp}{(2\pi\hbar)^d} \mathrm{d}^dk \left(\delta(\epsilon-\mu) + \frac{\pi^2T^2}{6}\delta^{\prime\prime}(\epsilon-\mu) + \cdots \right) k^2 |V_{\mathrm{imp}}(k)|^2 \delta(\vec k \cdot \vec v(p)) \notag \\
&= \frac{1}{n^2e^2}\int \frac{\mathrm{d}^dp}{(2\pi\hbar)^d} \mathrm{d}^dk \left(\delta(\epsilon-\mu) + \frac{\pi^2T^2}{6}\delta^{\prime\prime}(\epsilon-\mu) + \cdots \right) \frac{k^{d+1} |V_{\mathrm{imp}}(k)|^2}{v_{\mathrm{F}}} \notag \\
&\propto \frac{\delta^2}{n^2e^2\xi }\int \mathrm{d}\epsilon  \frac{\nu(\epsilon)}{v_{\mathrm{F}}(\epsilon)}\left(\delta(\epsilon-\mu) + \frac{\pi^2T^2}{6}\delta^{\prime\prime}(\epsilon-\mu) + \cdots \right) \notag \\
&\propto \rho(0) \left[1+ \frac{\pi^2T^2}{6} \left[\left(\frac{\nu(\mu)}{v_{\mathrm{F}}(\mu)}\right)^{\prime\prime} - \left(\frac{\nu(\mu)}{v_{\mathrm{F}}(\mu)}\right)^{\prime} \frac{\nu^\prime(\mu)}{\nu(\mu)}\right]\right]. 
\end{align}
In the last step, we have used the fact that $\mu$ is $T$ dependent, in order to keep $n$ fixed.
Depending on the band structure and $\mu$, this is a perturbative correction which does not, a priori, have a fixed sign.   The important point for us is that (working with a Fermi liquid where $1/\ell_\text{ee} \sim T^2$) \begin{equation}
\frac{\rho(T)-\rho(0)}{\rho(0)} \propto \left(\frac{T}{\mu}\right)^2 \propto \frac{\lambda_{\mathrm{F}}}{\ell_{\mathrm{ee}}}. \label{eq:lambdaFl}
\end{equation}
These temperature dependent corrections are significantly smaller than those caused by electron-electron interactions.  As we will see, the parameter governing the magnitude of corrections due to electron-electron interactions is $\xi/\ell_{\mathrm{ee}}$, which is much larger than (\ref{eq:lambdaFl}).

\subsection{Kinetic Theory on a 2d Fermi Surface}
\label{sec:toy}
\subsubsection{The Toy Model}
\label{sec:introtoy}
We now turn to a series of toy models of 2d Fermi liquids with circular Fermi surfaces of Fermi velocity $v_{\mathrm{F}}$, following the recent papers \cite{levitov2, 2016arXiv161200856L, levitov3}. The technical virtue of this model will be that only finite dimensional matrices need to be inverted to compute $\mathcal{A}(k)$ in (\ref{eq:sigmaexact4}), and hence the resistivity can be obtained exactly.  The model assumes that in the low temperature limit, the only interesting dynamics is associated with fluctuations in $f$ exactly at the Fermi surface, and so neglect all thermal effects.   While this is quite a strong assumption, it appears to model experiments of flows through tight constrictions reasonably well \cite{Molenkamp95,Moll1061,2017arXiv170306672K}.   To be more specific, we approximate the distribution function by
\begin{equation}\label{eq:jmodes}
\Phi(\vec x, \vec p) \approx \sum_{j\in\mathbb{Z}} \mathrm{e}^{\mathrm{i}jp_\theta} \Phi_j(\vec x),
\end{equation}
where $\tan p_\theta = p_y/p_x$ is the angle of the momentum vector.  More formally, we write \begin{equation}
|\Phi\rangle = \int \mathrm{d}^2x \sum_j \Phi_j(\vec x) |j(\vec x)\rangle, \;\;\;\; \text{ with }  |j(\vec x)\rangle \equiv \int \mathrm{d}^2p \; \frac{(p_x+\mathrm{i}p_y)^j}{p_{\mathrm{F}}^j}.
\end{equation}  
Henceforth we will denote $p_\theta$ as $\theta$ for simplicity.

Our first goal is to project $\mathsf{W}$ and $\mathsf{L}$ onto only the harmonic modes, labeled by $j$ in (\ref{eq:jmodes});  we follow the presentation of \cite{2016arXiv161200856L}.  As $T\rightarrow 0$, we anticipate that the only interesting dynamics occurs at the Fermi surface.   The fluctuations of the local number density of electrons are given by \begin{equation}
\mathrm{\Delta}n(\vec x) = \int \frac{\mathrm{d}^2p}{(2\pi\hbar)^2} \left(-\frac{\partial f_{\mathrm{eq}}}{\partial \epsilon}\right)\Phi = \nu(\mu)\Phi_0.
\end{equation} 
The momentum density $\mathfrak{g}_i$ is similarly given by \begin{subequations}\begin{align}
    \mathfrak{g}_x(\vec x) &= \int \frac{\mathrm{d}^2p}{(2\pi\hbar)^2} \left(-\frac{\partial f_{\mathrm{eq}}}{\partial \epsilon}\right) \Phi \, p_{\mathrm{F}}\cos\theta   = \nu(\mu)p_{\mathrm{F}} \frac{\Phi_1  +\Phi_{-1}}{2}, \\
    \mathfrak{g}_y(\vec x) &= \int \frac{\mathrm{d}^2p}{(2\pi\hbar)^2} \left(-\frac{\partial f_{\mathrm{eq}}}{\partial \epsilon}\right)\Phi \, p_{\mathrm{F}}\sin\theta  = \nu(\mu)p_{\mathrm{F}} \frac{\Phi_{-1} - \Phi_{1}}{2\mathrm{i}}.
\end{align}\end{subequations}
We conclude that $\Phi_{0,\pm 1}$ must be exactly conserved in the clean theory.   Denoting with $|j\rangle$ the mode $\Phi \propto \mathrm{e}^{\mathrm{i}j\theta}$, we conclude that the simplest non-trivial $\mathsf{W}$ respecting the conservation of charge and momentum is \begin{equation}
\mathsf{W} =\frac{1}{\nu(\mu)} \frac{v_{\mathrm{F}}}{\ell_{\mathrm{ee}}} \left({\mathds{1}}-|-1\rangle\langle -1|-|0\rangle\langle0| - |1\rangle\langle 1|\right) =
\frac{1}{\nu(\mu)} \frac{v_{\mathrm{F}}}{\ell_{\mathrm{ee}}} \sum_{|j|\geq 2} |j\rangle\langle j| .  \label{eq:toymodelW}
\end{equation}
$\ell_{\mathrm{ee}}$ is the mean free path for electron-electron momentum-conserving collisions:  it is the length scale over which higher harmonics in (\ref{eq:jmodes}) decay in the absence of any disorder.   Projected on to the Fermi surface harmonics, in the homogeneous theory: \begin{equation}\label{eq:Lsincos}
\mathsf{L} = \frac{\mathrm{i}v_{\mathrm{F}}}{\nu(\mu)} \left(\cos\theta k_x + \sin\theta k_y\right).
\end{equation}
This can also be transformed into the $|j\rangle$ basis, but it is not instructive to do so now.   Note that the factors of $1/\nu$ in both $\mathsf{W}$ and $\mathsf{L}$ are related to the non-trivial inner product (\ref{eq:innerproduct}), which in this toy model is relatively simple: \begin{equation}
\langle j|j^\prime\rangle = \nu(\mu)\mdelta_{jj^\prime}.  \label{eq:toymodelinner}
\end{equation}


\subsubsection{A Single Fermi Surface}
\label{sec:1FS}
In order to compute the resistivity we simply need to evaluate $\mathcal{A}(\vec k)$, given in (\ref{eq:sigmaexact4}).  Note that what was previously denoted as $|\mathsf{n}\rangle$  is now denoted as $|0\rangle$ -- the zeroth harmonic on the Fermi surface.   Hence, we must compute $\langle 0|(\mathsf{W}+\mathsf{L}(\vec k))^{-1}|0\rangle$.   We outline the computation in Appendix \ref{app:toy1}; it is quite similar to \cite{levitov2,2016arXiv161200856L,levitov3}  The result is \begin{equation}
\mathcal{A}(k) = \nu(\mu)\frac{\sqrt{1+k^2\ell^2_{\mathrm{ee}}}-1}{v_{\mathrm{F}}\ell_{\mathrm{ee}}k^2}. \label{eq:432}
\end{equation}
An immediate consequence of (\ref{eq:432}) is that for arbitrary (isotropic) inhomogeneity, electron-electron interactions in this model decrease the resistance.

It is instructive to consider first the limit where $\ell_{\mathrm{ee}}\gg \xi$, so that interactions are very weak.  In this case, the ``typical"  $k\sim 1/\xi$ and $k\ell_{\mathrm{ee}}$ is large;  in this limit we obtain \begin{equation}
\mathcal{A}(k) = \frac{\nu(\mu) }{v_{\mathrm{F}}k},
\end{equation}
and hence \begin{equation}
\rho_{xx} \propto \frac{\nu(\mu) \delta^2}{2n^2e^2 v_{\mathrm{F}}\xi}.
\end{equation}
Keeping track of constant prefactors more carefully, one can show exact agreement with the free theory result (\ref{eq:42result}).

In the opposite limit where $\xi \gg \ell_{\mathrm{ee}}$, we instead find \begin{equation}\label{eq:visclimit}
\rho_{xx}  = \frac{\nu(\mu)}{2n^2e^2} \int \frac{\mathrm{d}^2k}{(2\pi)^2} |V_{\mathrm{imp}}(k)|^2 \frac{\ell_{\mathrm{ee}}k^2}{2v_{\mathrm{F}}} = \frac{\nu \ell_{\mathrm{ee}}}{4v_{\mathrm{F}}n^2e^2}\mathbb{E}\left[ (\nabla V_{\mathrm{imp}})^2\right] = \eta \, \mathbb{E}\left[\left(\nabla \frac{1}{n}\right)^2\right],
\end{equation}
where we have used the fact that $\mathrm{\Delta} n \approx \nu V_{\mathrm{imp}}$ to leading order in $V_{\mathrm{imp}}$, along with the definition of the shear viscosity $\eta = n^2\ell_{\mathrm{ee}}/4\nu v_{\mathrm{F}}$ \cite{2016arXiv161200856L} in the last step.    \blue{The last equation above, expressing the resistivity in terms of the viscosity, was found in \cite{KS11}.}   We can interpret this last result from a perturbative hydrodynamic transport bound \cite{Lucas:2015lna}: when the only dissipative coefficient is shear viscosity we have $\rho_{xx} J_x^2  = \frac{\eta}{2}\mathbb{E}[(\partial_i v_j + \partial_j v_i - \partial_k v_k \delta_{ij})^2 ]$, on the function $v_x = J/n$.   Hence, in this regime we have transitioned from ballistic to hydrodynamic transport.

\begin{figure}
\centering
\includegraphics[width=3.5in]{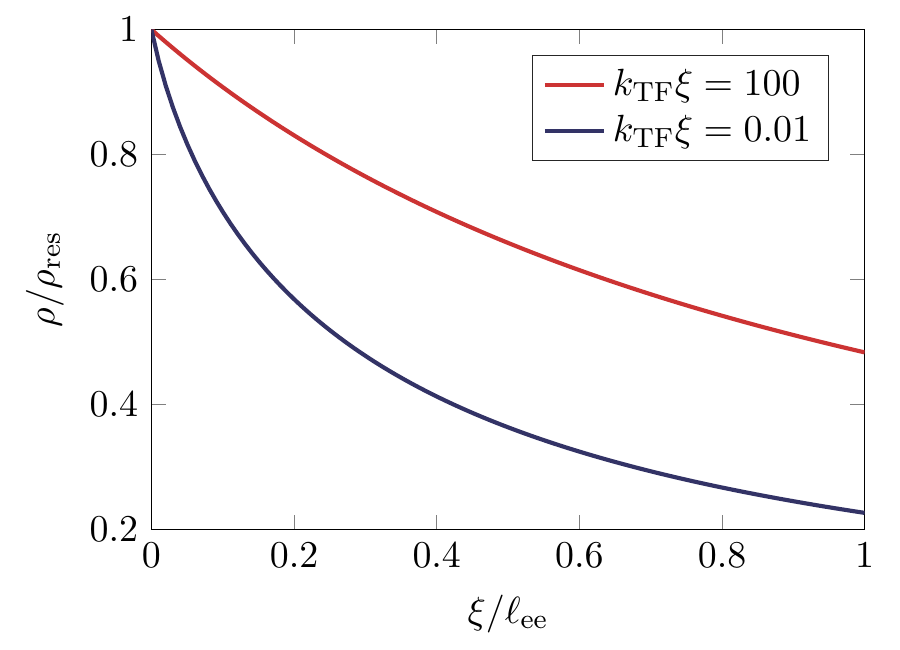}
\caption{The resistivity (\ref{eq:432}) as a function of the interaction strength $\xi/\ell_{\mathrm{ee}}$, measured relative to $\rho_{\mathrm{res}}$, the residual resistivity when $\ell_{\mathrm{ee}}=\infty$.   We have used the impurity potential (\ref{eq:kTF}).  Electron-electron interactions decrease the resistance, as it must in this toy model.  Furthermore, when $|V_{\mathrm{imp}}(k)|$ has significant weight at small $k$,  the effects of interactions is important even when $\ell_{\mathrm{ee}}\sim 10\xi$.}
\label{fig:432}
\end{figure}

Of course, through the kinetic theory solution, we in fact understand the entire crossover between these two regimes.  It is instructive to consider a specific form for $V_{\mathrm{imp}}(k)$, associated with random point-charge impurities, placed a distance $\xi$ above the 2d plane:  \begin{equation}
|V_{\mathrm{imp}}(\vec k)|^2 \propto \frac{\mathrm{e}^{-2|\vec k|\xi}}{(|\vec k|+k_{\mathrm{TF}})^2} \,. \label{eq:kTF}
\end{equation}
Here $k_{\mathrm{TF}}$ is a Thomas-Fermi screening wave number.  We cannot perform the integration in (\ref{eq:432}) analytically, but it is straightforward to do numerically.  The result is shown in Figure \ref{fig:432}.   As anticipated in (\ref{eq:visclimit}), as $\ell_{\mathrm{ee}}$ becomes shorter the resistivity decreases and ultimately tends to zero.  In the limit where $k_{\mathrm{TF}}\rightarrow 0$, the decrease in the resistivity is significantly faster due to the enhancement of $V_{\mathrm{imp}}(k)$ in the $k\rightarrow 0$ limit.

\subsubsection{Long-Lived $j=2$ Mode}
\label{sec:IN}

Now consider a slight twist to the previous model:  let us suppose that the $j=2$ `d-wave' modes are also long-lived relative to generic excitations. This will be seen to dramatically change the physics.  
We modify $\mathsf{W}$ to \begin{equation}
\mathsf{W} =  \frac{v_{\mathrm{F}}}{\ell_{\mathrm{ee}}} \left({\mathds{1}}-|-1\rangle\langle -1|-|0\rangle\langle0| - |1\rangle\langle 1| -(1-b)\left(|2\rangle\langle 2| + |-2\rangle\langle -2|\right)\right).
\end{equation}
The parameter $0<b<1$ determines the lifetime of the $j=2$ modes, relative to the higher harmonics:  they are exactly conserved if $b=0$;  when $b=1$, we recover the results of the previous subsection.   Following the techniques of Appendix \ref{app:toy1}, together with the calculation when $b=1$, leads to an analytic expression for $\mathcal{A}(k)$, and hence $\rho_{xx}$: \begin{equation}
\mathcal{A}(k) = \frac{\nu(\mu)}{v_{\mathrm{F}}}\frac{\ell_{\mathrm{ee}}}{\sqrt{1+k^2\ell^2_{\mathrm{ee}}}+2b-1}.  \label{eq:433}
\end{equation}

\begin{figure}
\centering
\includegraphics[width=3.5in]{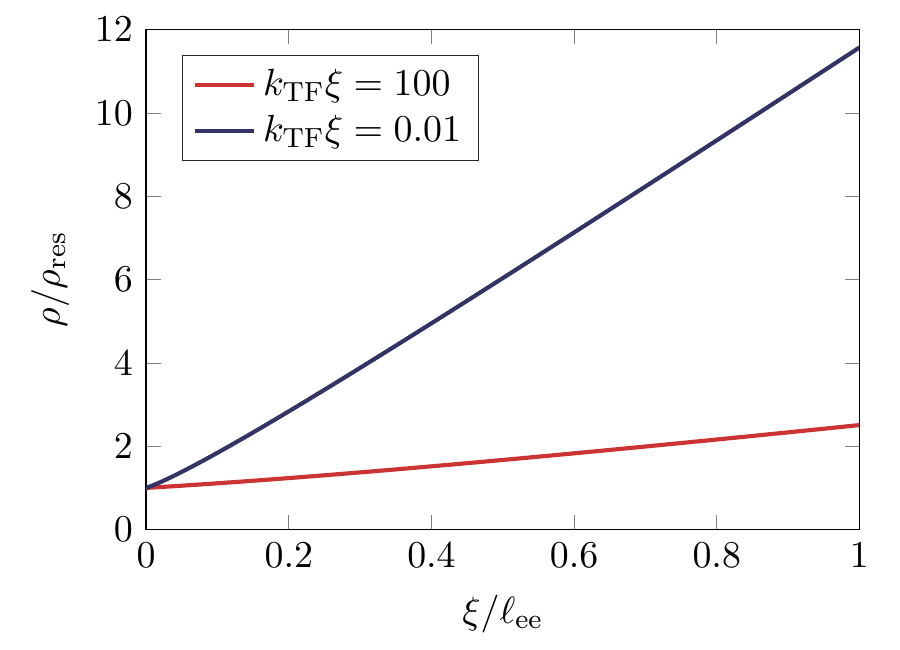}
\caption{The resistivity (\ref{eq:433}) as a function of the interaction strength $\xi/\ell_{\mathrm{ee}}$, measured relative to $\rho_{\mathrm{res}}$, the residual resistivity when $\ell_{\mathrm{ee}}=\infty$.   We have used the impurity potential (\ref{eq:kTF}).  Electron-electron interactions enhance the resistance when $b=0$, as they must in this toy model.  As in Figure \ref{fig:432}, we observe that the effects of interactions can become important even when $\ell_{\mathrm{ee}} \sim 10\xi$ if the disorder is sufficiently correlated on long length scales.}
\label{fig:433}
\end{figure}

Let us begin by setting $b=0$ -- in this case the $j=2$ mode is exactly conserved.  It is not difficult to see in this case that because \begin{equation}
k < \frac{k^2\ell_{\mathrm{ee}}}{\sqrt{1+k^2\ell^2_{\mathrm{ee}}}-1}, \label{eq:b0k}
\end{equation}
electron-electron interactions strictly enhance the resistivity.  Moreover, when interactions are strong so that $\xi \gg \ell_\text{ee}$, \begin{equation}
\rho_{xx} \approx \frac{\nu(\mu)}{2n^2e^2v_{\mathrm{F}}} \int \frac{k\mathrm{d}k}{2\pi} |V_{\mathrm{imp}}|^2 \frac{2}{\ell_{\mathrm{ee}}} = \frac{\nu(\mu)\delta^2}{n^2e^2v_{\mathrm{F}}\ell_{\mathrm{ee}}}.  \label{eq:433b0}
\end{equation}
We observe that, up to the small factor of $\delta^2$, it is as if the momentum-relaxing rate is actually set by $\ell_{\mathrm{ee}}$ -- the mean free path for momentum-conserving collisions.  We will see how this, potentially counter-intuitive, effect can be understood from general principles in Section \ref{sec:genprin}.   Figure \ref{fig:433} shows this effect numerically.    When $V_{\mathrm{imp}}$ is associated with out-of-plane point charges, and screening is weak, we observe a remarkable effect: an accidental approximate Mattheisen rule which holds well across the entire ballistic-to-hydrodynamic crossover, so that  \begin{equation}
\rho_{xx} \approx \frac{c_1}{\xi} + \frac{c_2}{\ell_{\mathrm{ee}}}, \;\;\; c_{1,2} \text{ constants}.
\end{equation}

Let us also note that in this case with $b=0$, the contribution to the resistivity due to long-wavelength correlations in the impurity potential when $V_{\mathrm{imp}}$ is given by (\ref{eq:kTF}) is  \begin{equation}
\rho_{xx} \propto \int \frac{k\mathrm{d}k}{2\pi} \frac{2\mathrm{e}^{-2k\xi}}{(k+k_{\mathrm{TF}})^2} \left(1+\mathrm{O}\left((k\ell_{\mathrm{ee}})^2\right)\right).
\end{equation}
If $k_{\mathrm{TF}} \rightarrow 0$, this integral is logarithmically divergent:  schematically, \begin{equation}
\rho_{xx} \propto \log \frac{1}{k_{\mathrm{TF}}\max(\xi,\ell_{\mathrm{ee}})}.
\end{equation}
Although it may well be the case that such divergences are cured at higher orders in perturbation theory, materials with low screening ($k_{\mathrm{TF}}\rightarrow 0$) would be expected to have an extremely high resistivity.  The effects of weak disorder could be compensated by this logarithmic enhancement of $\rho_{xx}$ to provide a momentum relaxation length comparable to $\ell_{\mathrm{ee}}$.

\begin{figure}[h]
\centering
\includegraphics[width=6in]{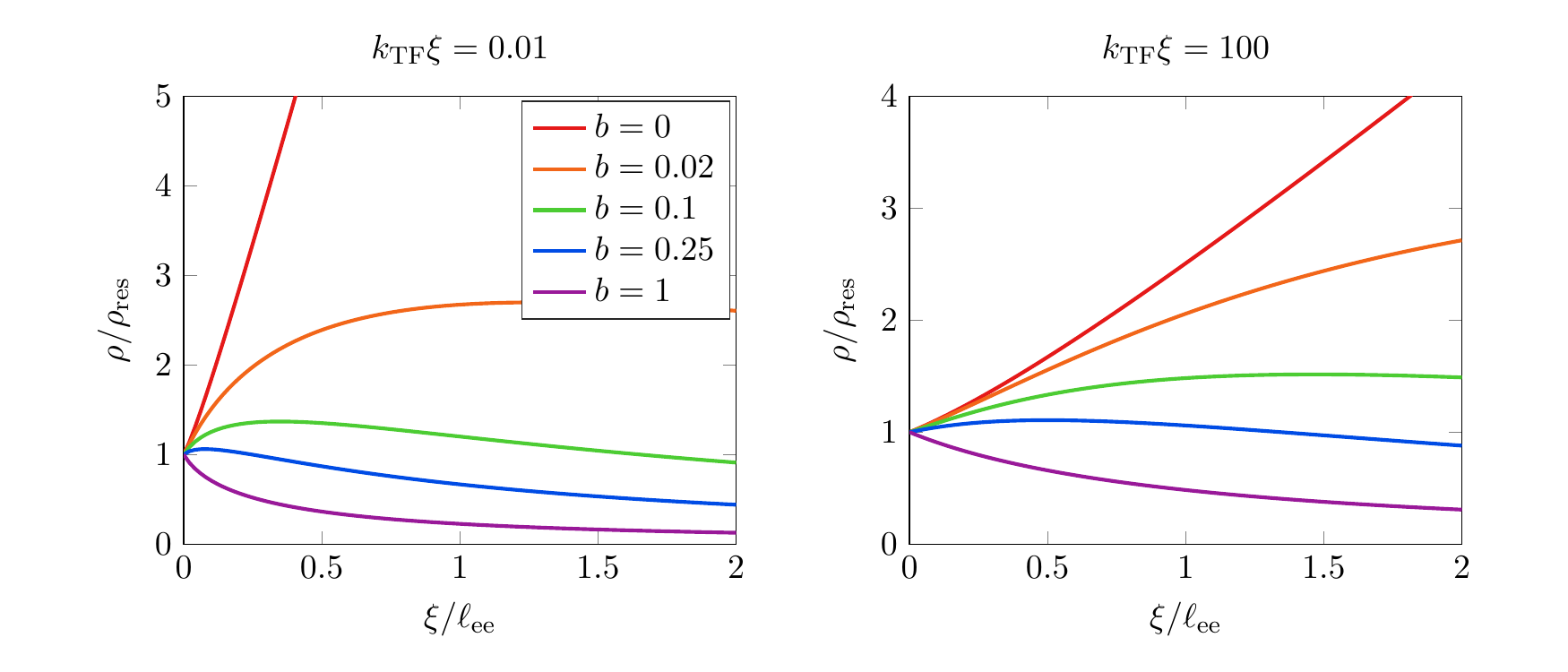}
\caption{The resistivity (\ref{eq:433}) as a function of the interaction strength $\xi/\ell_{\mathrm{ee}}$, for different values of $b$, measured relative to $\rho_{\mathrm{res}}$, the residual resistivity when $\ell_{\mathrm{ee}}=\infty$.   We have used the impurity potential (\ref{eq:kTF}).  As $b$ becomes smaller, we observe a larger regime where electron interactions enhance the resistivity.}
\label{fig:433cross}
\end{figure}

We now turn to the case $0<b<1$.   Following the discussion of (\ref{eq:b0k}), from (\ref{eq:433}) it is not difficult to see that the resistivity can only be enhanced by interactions when $b<\frac{1}{2}$.   When $b\ll 1$ is parametrically small, and $\ell_{\mathrm{ee}}\ll \xi$, it is helpful to approximate (\ref{eq:433}) by \begin{equation}
\rho_{xx} \approx \frac{\nu(\mu)}{2n^2e^2v_{\mathrm{F}}} \int \frac{k\mathrm{d}k}{2\pi} |V_{\mathrm{imp}}|^2 \frac{2k^2\ell_{\mathrm{ee}}}{k^2\ell^2_{\mathrm{ee}}+4b}.
\end{equation}
Whenever $\ell_{\mathrm{ee}}/\xi \gtrsim 2\sqrt{b}$, the resistivity will be well-approximated by the $b=0$ limit given in (\ref{eq:433b0}):  interactions strictly enhance the conductivity.  In contrast, whenever $\ell_{\mathrm{ee}}/\xi \lesssim 2\sqrt{b}$, then we find \begin{equation}
\rho_{xx}\approx \frac{1}{2b}\times \frac{\nu \ell_{\mathrm{ee}}}{4v_{\mathrm{F}}n^2e^2}\mathbb{E}\left[ (\nabla V_{\mathrm{imp}})^2\right]  = \eta(b) \mathbb{E}\left[\left(\nabla \frac{1}{n}\right)^2\right],
\end{equation}
with $\eta(b)\approx \eta(1)/2b$ for $b \ll 1$.   So long as $b>0$, therefore, for strong enough interactions we do ultimately recover a more conventional viscous-dominated hydrodynamic regime, analogous to Section \ref{sec:1FS}.   However, when $b$ is small, there may be a parametrically large regime where the long-lived $j=2$ mode enhances the resistivity.  We explore this crossover numerically in Figure \ref{fig:433cross}.  Especially when the disorder potential has a significant long-wavelength component, we observe that the viscous regime where interactions suppress the resistivity can emerge extremely quickly.

\subsubsection{Two Fermi Surfaces: Baber Scattering, Revisited}
\label{sec:2FS}

Another complication of our original model is to consider the case where the Fermi surface consists of two disconnected pockets of circular shape.  If the pockets are sufficiently well separated in the Brillouin zone, with the distance between pockets on the scale of $k_{\mathrm{F}}$, and the Brillouin zone is sufficiently large, then we may neglect both inter-pocket scattering of electrons and umklapp processes.   In this limit, the number of electrons within each pocket must be conserved separately. \blue{We refer to the difference in electron density between the two pockets as an imbalance mode.}

The simplest model for this consists of two copies of our model of a single Fermi surface, with an additional collision term that can exchange momentum between the two pockets (as 2-body scattering events may allow electrons from one pocket to dissipate momentum into the other).   Let us denote with $\Phi^A$ ($A=1,2$) the angular distribution function in a single pocket, and with $|jA\rangle$ the $j^{\mathrm{th}}$ harmonic of $\Phi^A$.   The notational simplifications from Section \ref{sec:introtoy} carry through otherwise.  

Let us begin by assuming, for ease of computation, that the electrons in each pocket have identical $\nu(\mu)$ and \blue{quadratic disperion relation $\epsilon(p) \propto p^2-p_{\mathrm{F}}^2$}, and that each pocket of the Fermi surface is circular.  In this case, we write the streaming term as \begin{equation}
\mathsf{L} = \mathrm{i} \left(\cos\theta k_x + \sin\theta k_y\right) \left(v_{\mathrm{F,1}} \mathsf{P}_1 + v_{\mathrm{F,2}} \mathsf{P}_2
\right).  \label{eq:L434}
\end{equation}
$\mathsf{P}_{1}$ is a projection matrix, defined such that $\mathsf{P}_1|jA\rangle = |j1\rangle \delta_{A1}$; $\mathsf{P}_2$ is defined similarly.
We write the collision term as
\begin{equation}
\mathsf{W} = \frac{v_{\mathrm{F,2}}}{\ell_{\mathrm{ee}}}\sum_{|j|\ge 2, A} |jA\rangle\langle jA| + \frac{v_{\mathrm{F,2}}}{\ell_{\mathrm{ee}}}\sum_{j=\pm 1} \frac{(v_{\mathrm{F,2}} |j1\rangle - v_{\mathrm{F,1}} |j2\rangle) ( v_{\mathrm{F,2}} \langle j1| -  v_{\mathrm{F,1}} \langle j2|) }{v_{\mathrm{F,1}}^2+v_{\mathrm{F,2}}^2},
\end{equation}  
For simplicity we have taken the relaxation time $\tau_\text{ee} = \ell_\text{ee}/v_{F,2}$ to be the same for all non-conserved modes.  Our qualitative results are not sensitive to this assumption.
The second term accounts for the fact that only the total momentum \begin{equation}
    |p_{\pm}\rangle \propto \frac{v_{\mathrm{F,1}}}{\sqrt{v_{\mathrm{F,1}}^2+v_{\mathrm{F,2}}^2}}|\pm1, 1\rangle + \frac{v_{\mathrm{F,2}}}{\sqrt{v_{\mathrm{F,1}}^2+v_{\mathrm{F,2}}^2}}|\pm1, 2\rangle 
\end{equation} is conserved.  \blue{The momentum vector takes this form because with a quadratic dispersion relation, the velocity and momentum are proportional.}  For simplicity, we assume that electrons in both pockets have the same collision rate.  The global density is given by \begin{equation}
|\mathsf{n}\rangle \propto  |01\rangle +  |02\rangle.
\end{equation} 

\begin{figure}
\centering
\includegraphics[width=6.5in]{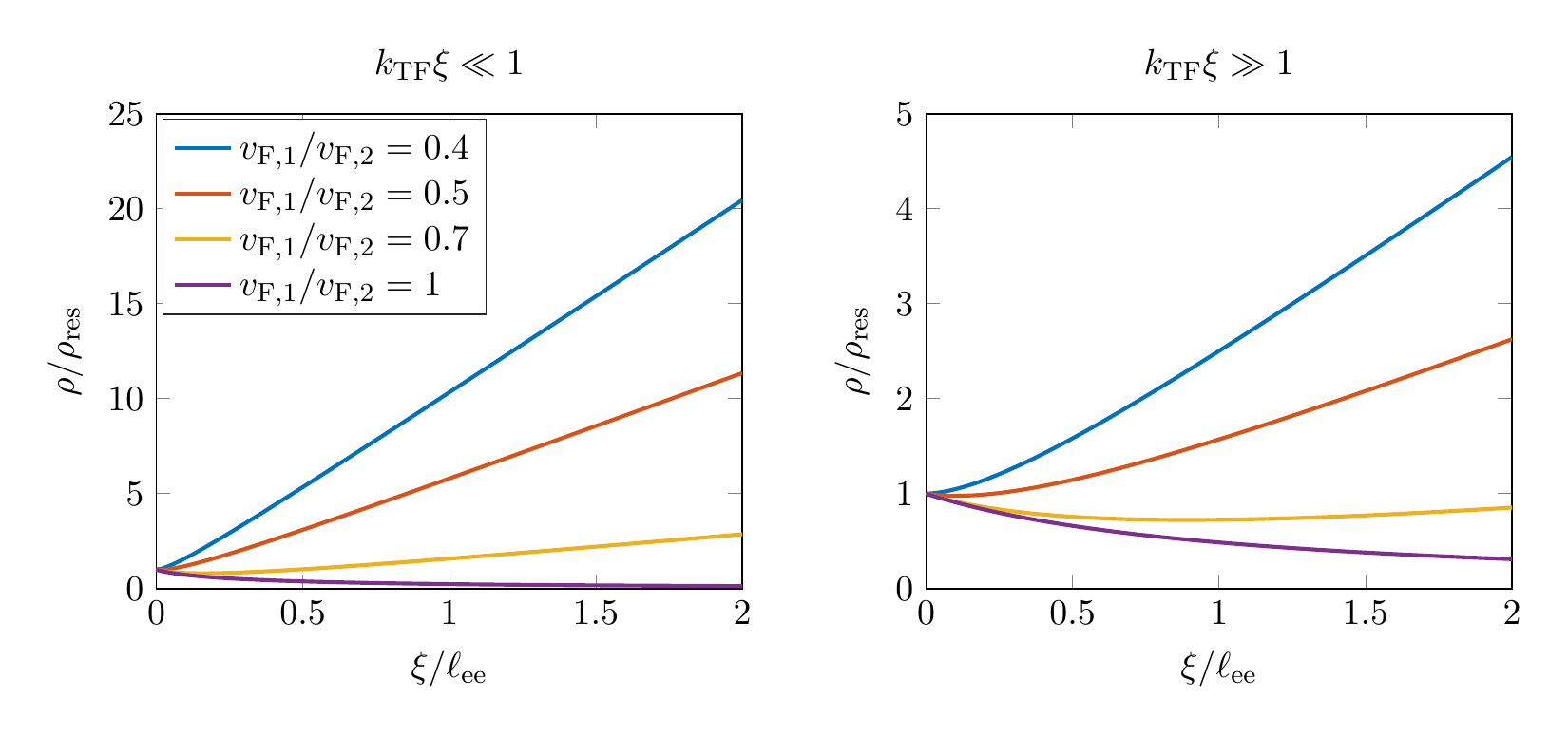}
\caption{The resistivity as a function of the interaction strength $\xi/\ell_{\mathrm{ee}}$, for different values of $v_{\mathrm{F,1}}/v_{\mathrm{F,2}}$, measured relative to $\rho_{\mathrm{res}}$, the residual resistivity when $\ell_{\mathrm{ee}}=\infty$.   We have used the impurity potential (\ref{eq:kTF}).  As the particles in each pocket move at different velocities, we observe a rapid enhancement of the resistivity, with $\rho \sim 1/\ell_\text{ee}$.}
\label{fig:baber}
\end{figure}

While it is challenging to analytically compute $\mathcal{A}(\vec k)$, using the techniques of Appendix \ref{app:toy1} it is straightforward to compute $\mathcal{A}(\vec k)$, and hence $\rho$, numerically.   The result is shown in Figure \ref{fig:baber}.  As usual, we have rescaled the results relative to the residual resistivity at $\ell_{\mathrm{ee}}=\infty$.   We observe that when $\ell_{\mathrm{ee}}$ is large, but finite, interactions decrease the resistivity.  This is due to viscous effects within each band.   As $\ell_{\mathrm{ee}}$ becomes small,  if the velocities of the two bands are not equivalent, then the fluctuations in the chemical potential help to source an ``imbalance mode".    This is precisely the mechanism that we argued in Section \ref{sec:summary} would lead to $\rho \propto \ell_{\mathrm{ee}}^{-1}$.

\begin{figure}
\centering
\includegraphics[width=3.8in]{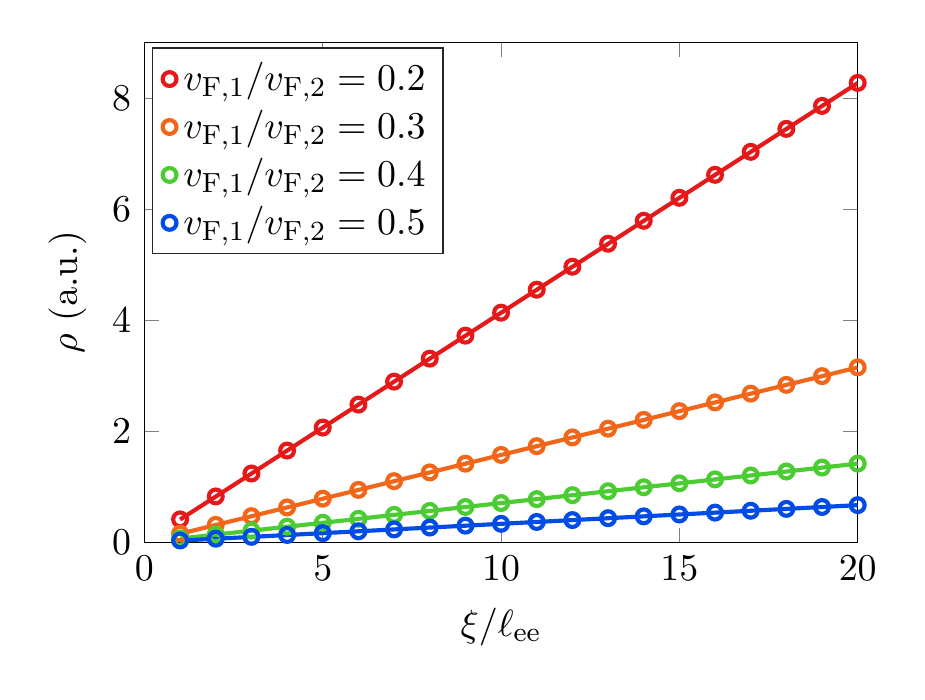}
\caption{The resistivity of the two-band Fermi liquid is dominated by imbalance diffusion in the hydrodynamic regime, and we see excellent quantitative agreement between our analytic prediction (\ref{eq:434final}) (solid lines) and numerical data (circles) in the hydrodynamic regime $\xi \gtrsim \ell_{\mathrm{ee}}$.  We have explicitly computed the coefficient of proportionality in (\ref{eq:434final}) from the form of $|V_{\mathrm{imp}}^2|$; there are no fit parameters in the comparison between numerics and analytics.}
\label{fig:434check}
\end{figure}

\blue{To be more quantitative, we observe that the hydrodynamic description of transport in this Fermi liquid contains an imbalance mode \cite{ushydro}. This mode can be seen already in the homogeneous system, with no disorder. Let us focus on flows in the long wavelength limit where $\ell_{\mathrm{ee}}$ is very short compared to the length scales over which $\Phi$ changes. We can choose to look at flows which only depend on the $x$ direction.  Taking the inner product of the Boltzmann equation (\ref{eq:LAmain}) -- in the absence of an external electric field -- with $j\le 2$ harmonics, we obtain
\begin{subequations}\begin{align}
 0 &=  \partial_x \Phi^1_1 = \partial_x \Phi^2_1, \\
 0 &=  \frac{v_{\mathrm{F,1}}}{v_{\mathrm{F,2}}}\partial_x \left(\Phi^1_0+\Phi^1_2\right) + \frac{2v_{\mathrm{F,2}}}{(v_{\mathrm{F,1}}^2 + v_{\mathrm{F,2}}^2)\ell_{\mathrm{ee}}} \left(v_{\mathrm{F,2}}\Phi^1_1-v_{\mathrm{F,1}}\Phi^2_1\right) \,, \\
  0 &=  \partial_x \left(\Phi^2_0+\Phi^2_2\right) - \frac{2v_{\mathrm{F,1}}}{(v_{\mathrm{F,1}}^2 + v_{\mathrm{F,2}}^2)\ell_{\mathrm{ee}}} \left(v_{\mathrm{F,2}}\Phi^1_1-v_{\mathrm{F,1}}\Phi^2_1\right), \\
  0 &= \frac{v_{\mathrm{F,1}}}{v_{\mathrm{F,2}}}\partial_x \Phi^1_1 + \frac{2\Phi^1_2}{\ell_{\mathrm{ee}}}, \\
    0 &= \partial_x \Phi^2_1 + \frac{2\Phi^2_2}{\ell_{\mathrm{ee}}} \,.
\end{align}\end{subequations}
In the above equations we have denoted $\Phi^A_j = \langle j,A|\Phi\rangle$, and we have used the fact that on flows with only $x$-dependence,  $\Phi^A_j = \Phi^A_{-j}$, to simplify the equations slightly.   These equations follow from (\ref{eq:LAmain}); following \cite{2016arXiv161200856L} we have dropped $j\ge 3$ harmonics in both bands as these are parametrically small at long wavelength and can be neglected.   Keeping the terms with fewest derivatives, and defining \begin{equation}
    v_x = v_{\mathrm{F,1}}\Phi^1_1 +v_{\mathrm{F,2}}\Phi^2_1  \,,
\end{equation}
we find that the time-independent hydrodynamic equations for this two-band model are \begin{subequations}\label{eq:2bandhydro}
\begin{align}
0 &\approx \partial_x \left( v_{\mathrm{F,1}}^2 v_x -\frac{1}{2}v_{\mathrm{F,1}}^2v_{\mathrm{F,2}}\ell_{\mathrm{ee}} \partial_x \left(\Phi^1_0 - \Phi^2_0\right)\right), \\ 
0 &\approx \partial_x \left( v_{\mathrm{F,2}}^2 v_x -\frac{1}{2}v_{\mathrm{F,1}}^2v_{\mathrm{F,2}}\ell_{\mathrm{ee}} \partial_x \left(\Phi^2_0 - \Phi^1_0\right)\right), \\ 
0 &\approx \partial_x \left(v_{\mathrm{F,1}}^2 \Phi_0^1 + v_{\mathrm{F,2}}^2 \Phi_0^2\right) - \ell_{\mathrm{ee}} \frac{v_{\mathrm{F,1}}^4 + v_{\mathrm{F,2}}^4}{2v_{\mathrm{F,2}}(v_{\mathrm{F,1}}^2 + v_{\mathrm{F,2}}^2)} \partial_x^2 v_x.
\end{align}
\end{subequations}
We have thrown out terms that are third order or higher in derivatives, as they are subleading in the hydrodynamic limit.  These hydrodynamic equations make manifest that there is a diffusive imbalance mode with a diffusive current $\propto \partial_x(\Phi^1_0 - \Phi^2_0)$.  

Now, we turn to the computation of the resistivity.  In the hydrodynamic limit $k\ell_{\mathrm{ee}}\ll 1$, we can approximately compute $\mathcal{A}(k)$ analytically   by inverting the $10\times 10$ submatrix of $\mathsf{W}+\mathsf{L}$ consisting of only $j\le 2$ modes.  This correctly computes $\mathcal{A}(k)$ to leading order in $k$.  We find that so long as $v_{\mathrm{F,1}} \ne v_{\mathrm{F,2}}$: \begin{equation}
    \mathcal{A}(k) \approx \frac{2 \nu(\mu) \left(v_{\mathrm{F,1}}^2 - v_{\mathrm{F,2}}^2\right)^2}{\ell_{\mathrm{ee}}v_{\mathrm{F,2}}v_{\mathrm{F,1}}^2\left(v_{\mathrm{F,1}}^2 + v_{\mathrm{F,2}}^2\right)k^2}, 
\end{equation}
which, following our discussion in Section \ref{sec:IN}, implies that \begin{equation}
    \rho \sim \frac{v_{\mathrm{F,2}}}{\ell_\mathrm{ee}} \frac{2 \left(v_{\mathrm{F,1}}^2 - v_{\mathrm{F,2}}^2\right)^2}{v^2_{\mathrm{F,2}}v_{\mathrm{F,1}}^2\left(v_{\mathrm{F,1}}^2 + v_{\mathrm{F,2}}^2\right)}.  \label{eq:434final}
\end{equation}
The scattering length $\ell_\mathrm{ee}$ in (\ref{eq:434final}) is a direct consequence of the diffusivity proportional to $\ell_\mathrm{ee}$ in (\ref{eq:2bandhydro}). If this imbalance mode were absent, then we would obtain instead the viscous result ${\mathcal{A}} \sim \ell_\text{ee}$ as in section \ref{sec:1FS}. The velocity dependence of (\ref{eq:434final}) comes from the thermodynamic susceptibility characterizing the overlap between the charge density and the imbalance density \cite{ushydro}. In particular, for identical realizations of disorder, but differing values of $v_{\mathrm{F,1}}$ and $v_{\mathrm{F,2}}$, (\ref{eq:434final}) gives us a simple way to confirm that our numerically calculated resistivity is in the hydrodynamic regime and that the resistivity is dominated by imbalance diffusion: see Figure \ref{fig:434check}.
}

\blue{There are other mechanisms that can lead to imbalance modes.  For example, in charge-neutral graphene, the relativistic dispersion relation forbids electron-hole scattering at lowest order in interactions \cite{foster, schutt}.  In charge neutral graphene, the temperature dependence of thermodynamic susceptibilities is not negligible, and so this is not a good model system to observe $\rho \sim 1/\ell_{\mathrm{ee}}$.  }

\subsubsection{Toy Model of Electron-Phonon Scattering}
\label{sec:EPH}
\begin{figure}
\centering
\includegraphics[width=6.5in]{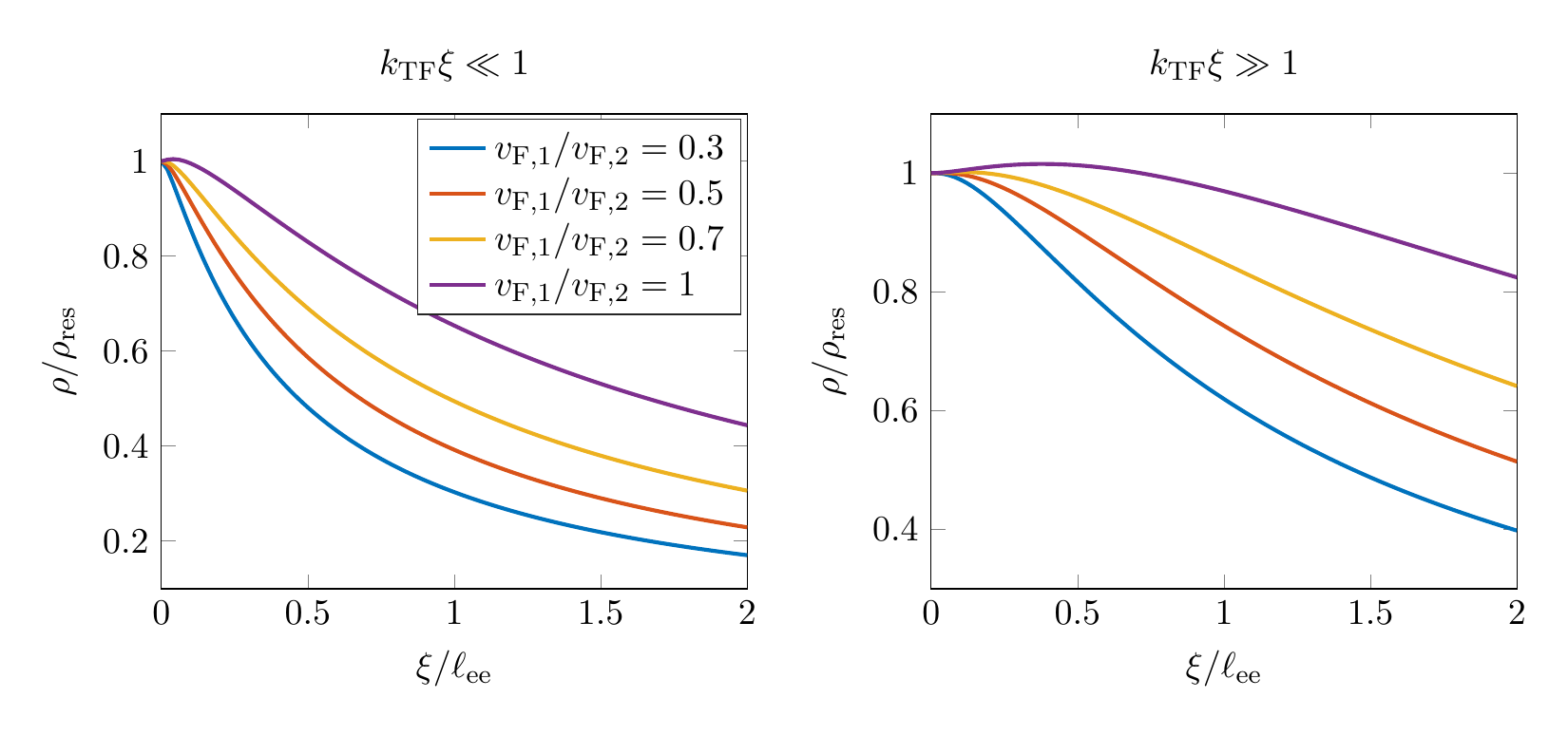}
\caption{The resistivity of the toy model of electrons and phonons, as a function of the interaction strength $\xi/\ell_{\mathrm{ee}}$, for different values of $v_{\mathrm{F,1}}/v_{\mathrm{F,2}}$, measured relative to $\rho_{\mathrm{res}}$, the residual resistivity when $\ell_{\mathrm{ee}}=\infty$.   We have used the impurity potential (\ref{eq:kTF}).   The physics is qualitatively identical to the viscous-dominated transport discussed in Section \ref{sec:1FS}.}
\label{fig:eph}
\end{figure}

Finally, let us briefly discuss a very crude toy model of electron-phonon scattering.   We use an identical model to Section \ref{sec:2FS}, thinking of band 1 as describing phonons, and band 2 as describing electrons.  The distribution $\Phi^1$ of phonons should no longer be interpreted as a Fermi surface, but simply as the total number of phonons with velocity at angle $\theta$. Because phonon number is not conserved, we now take the interaction to be:
\begin{equation}
\mathsf{W}  = \frac{v_{\mathrm{F,2}}}{\ell_{\mathrm{ee}}} |01\rangle\langle 01|+ \frac{v_{\mathrm{F,2}}}{\ell_{\mathrm{ee}}}\sum_{|j|\ge 2, A} |jA\rangle\langle jA| + \frac{v_{\mathrm{F,2}}}{\ell_{\mathrm{ee}}}\sum_{j=\pm 1} \frac{(v_{\mathrm{F,2}} |j1\rangle - v_{\mathrm{F,1}} |j2\rangle) ( v_{\mathrm{F,2}} \langle j1| -  v_{\mathrm{F,1}} \langle j2|) }{v_{\mathrm{F,1}}^2+v_{\mathrm{F,2}}^2}.
\end{equation}
Again for simplicity we have taken the decay rate $\tau_{\text{ee}} = \ell_{\text{ee}}/v_{\mathrm{F},2}$ to be the same for all non-conserved quantities.
The zero mode of the phonons, in particular, is no longer a conserved quantity.  Furthermore, the charge density mode $|\mathsf{n}\rangle \propto |02\rangle$, since only the electrons are charged.    These changes will destroy the diffusive mode in the generalized hydrodynamics of the electron-phonon system.  As such, interactions should ultimately decrease the resistivity, as there is no diffusive mode decoupled from momentum drag.   We confirm this with a numerical computation of $\rho$ in Figure \ref{fig:eph}.   We expect that a more detailed quantitative treatment of electron-phonon interactions will lead to the same qualitative effects.

\subsection{General Principles}
\label{sec:genprin}
Now that we have seen a variety of toy models, let us now describe, in general, the circumstances under which electron-electron interactions will increase or decrease the resistivity,  in the weak disorder limit.

We now analyze the formula (\ref{eq:sigmaexact4}) for $\mathcal{A}(\vec k)$ for general kinetic theories \blue{with inversion and time-reversal symmetry, and with momentum-relaxation arising only from charge impurities $V_{\mathrm{imp}}$}.  It is instructive to break $\mathsf{W}$ and $\mathsf{L}$ into block $4 \times 4$ matrices, keeping track of (\emph{i}) odd vs. even vectors under inversion symmetry, and (\emph{ii}) null vectors of $\mathsf{W}$ (`slow' modes) vs. non-null vectors of $\mathsf{W}$ (`fast' modes).      We write \begin{equation}
|\Phi\rangle = \left(\begin{array}{c} |\Phi_{\mathrm{even,slow}}\rangle \\ |\Phi_{\mathrm{odd,slow}}\rangle  \\  |\Phi_{\mathrm{even,fast}}\rangle \\ |\Phi_{\mathrm{odd,fast}}\rangle \end{array}\right), \;\;\;\;\; \mathsf{W}+\mathsf{L} = \left(\begin{array}{cccc} 0 &\ \mathsf{L}_{\mathrm{es,os}} &\ 0 &\ \mathsf{L}_{\mathrm{es,of}} \\ \mathsf{L}_{\mathrm{os,es}} &\ 0 &\ \mathsf{L}_{\mathrm{os,ef}} &\ 0 \\ 0 &\ \mathsf{L}_{\mathrm{ef,os}} &\ \mathsf{W}_{\mathrm{e}} &\ \mathsf{L}_{\mathrm{ef,of}} \\ \mathsf{L}_{\mathrm{of,es}} &\ 0 &\ \mathsf{L}_{\mathrm{of,ef}} &\ \mathsf{W}_{\mathrm{o}}  \end{array}\right) \,.  \label{eq:4x4}
\end{equation}
We remind the reader that in the analysis of (\ref{eq:sigmaexact4}), the matrix inverse $(\mathsf{W}+\mathsf{L})^{-1}$ is taken over momentum indices alone.  Hence,  $\mathsf{L} = \mathrm{i}\vec k \cdot \vec v(\vec p)$.   We also remind the reader that $\mathsf{L}_{\mathrm{e}a,\mathrm{o}b}^{\mathsf{T}} = -\mathsf{L}_{\mathrm{o}b,\mathrm{e}a}$ for $a,b\in \lbrace \mathrm{s,f}\rbrace$.


Because the density vector $|\mathsf{n}\rangle$ is in the even/slow sector, it is clear from (\ref{eq:sigmaexact4}) that we need to compute the slow/even diagonal block of $(\mathsf{W}+\mathsf{L})^{-1}$.   It is straightforward, but tedious, to use block matrix identities to perform this matrix inverse.   As the result is quite cumbersome, we present  explicit results in Appendix \ref{app:genprin}, only focusing on the qualitative physics here.  \blue{For simplicity, we assume that all scattering rates in $\mathsf{W}_{\mathrm{o}}$ and $\mathsf{W}_{\mathrm{e}}$ are comparable, and so the discussion below is not applicable to the model of Section \ref{sec:IN} when $0<b\ll 1$.}

\blue{The first step of evaluating (\ref{eq:sigmaexact4}) is to ``integrate out" the fast degrees of freedom, which leads to \begin{equation}
    (\mathsf{W}+\mathsf{L})^{-1}_{\mathrm{slow}} = \left(\begin{array}{cc} \widehat{\mathsf{W}}_{\mathrm{e}} &\  \widehat{\mathsf{L}}_{\mathrm{es,os}} \\ \widehat{\mathsf{L}}_{\mathrm{os,es}} &\ \widehat{\mathsf{W}}_{\mathrm{o}} \end{array}\right)^{-1} \,.  \label{eq:44top22}
\end{equation}
We have only displayed the top-left $2\times 2$ sub-matrix of $(\mathsf{W}+\mathsf{L})^{-1}$ here. Explicit formulae for the matrices $\widehat{\mathsf{W}}$ and $\widehat{\mathsf{L}}$ can be found in Appendix \ref{app:genprin}. Physically speaking, $\widehat{\mathsf{W}}_{\mathrm{e,o}}$ each give the decay rate of spatial fluctuations of conserved quantities, and so they will scale with $k$ as follows: \begin{equation}
    \widehat{\mathsf{W}}_{\mathrm{e,o}} \sim k \min(1,k\ell_{\mathrm{ee}}).  \label{eq:Weoscaling}
\end{equation}
Some explanation of this result is warranted.  At ballistic length scales $k\ell_{\mathrm{ee}}\gg 1$, the density of quasiparticles at every momentum is approximately conserved and approximately all of these quantities are relaxed by impurities, which occurs at a rate $\propto k$, determined by the time over which quasiparticles traverse the inhomogeneous landscape.    This result was demonstrated explicitly at $\ell_{\mathrm{e}}= \infty$ in Section \ref{sec:free} and it will be perturbatively corrected in $\ell_{\mathrm{ee}}^{-1}$.   In the hydrodynamic limit,  $\widehat{\mathsf{W}} \sim k^2\ell_{\mathrm{e}}$ because spatial inhomogeneities of conserved quantities relax via diffusion (even if there is ballistic sound motion at leading order).   Regardless of $\ell_{\mathrm{ee}}$, we find that \begin{equation}
    \widehat{\mathsf{L}}_{\mathrm{es,os}},\widehat{\mathsf{L}}_{\mathrm{os,es}} \sim k \label{eq:Leoscaling}
\end{equation} 
from the explicit formulae in the Appendix.

Assuming $\widehat{\mathsf{W}}_{\mathrm{o}}$ is invertible (we discuss the more general case in the appendix):
\begin{equation}
    \mathcal{A}(k) = \langle \mathsf{n}| \left(\widehat{\mathsf{W}}_{\mathrm{e}} + \widehat{\mathsf{L}}_{\mathrm{es,os}}\widehat{\mathsf{W}}_{\mathrm{o}}^{-1}\widehat{\mathsf{L}}_{\mathrm{es,os}}^{\mathsf{T}} \right)^{-1}|\mathsf{n}\rangle.  \label{eq:Ak44}
\end{equation}
In the ballistic limit, both terms are proportional to $k$ and so $\mathcal{A}\sim k^{-1}$ as in Section \ref{sec:free}.  In the hydrodynamic limit, the first term scales as $k^2\ell_{\mathrm{ee}}$ while the second scales as $\ell_{\mathrm{ee}}^{-1}k^0$.  Thus, we would generically expect the second term to dominate the matrix inverse, leading to \begin{equation}
    \mathcal{A}(k) \sim \ell_{\mathrm{ee}}.
\end{equation}
This is precisely the result we found in the viscous-dominated hydrodynamic limit in Section \ref{sec:1FS}, and leads to $\rho \propto \ell_{\mathrm{ee}}$.  In this case, interactions enhance transport and reduce the resistivity.

However, it may be the case that there are more even conserved quantities than odd conserved quantities.\footnote{For this count, momentum should be counted as 1 conserved quantity and not $d$.}  This was the case, for example, in the model of imbalance diffusion in Section \ref{sec:2FS}.  In this case, $\widehat{\mathsf{L}}_{\mathrm{es,os}}\widehat{\mathsf{W}}_{\mathrm{o}}^{-1}\widehat{\mathsf{L}}_{\mathrm{os,es}}^{\mathsf{T}}$ is not a full rank matrix and it cannot, by itself, be inverted.  We conclude that if $|\mathsf{n}\rangle$ is a ``generic" even conserved quantity, that $\mathcal{A}(k)$ will be dominated by the part of $|\mathsf{n}\rangle$ lying in the null space of  $\widehat{\mathsf{L}}_{\mathrm{es,os}}\widehat{\mathsf{W}}_{\mathrm{o}}^{-1}\widehat{\mathsf{L}}_{\mathrm{os,es}}^{\mathsf{T}}$.  This leads to \begin{equation}
    \mathcal{A}(k) \sim \frac{1}{\ell_{\mathrm{ee}}k^2}
\end{equation}
and hence, as in Section \ref{sec:2FS}, $\rho \propto 1/\ell_{\mathrm{ee}}$: interactions supress transport and enhance $\rho$.  

The constant prefactors that we have neglected in the above discussion can be straightforwardly accounted for but will be sensitive to the specific microscopic model.  Below, we will see that such coefficients admit a simple interpretation when $k\ell_{\mathrm{ee}} \ll 1$.
}

\blue{In the hydrodynamic limit, we can provide more intuition for these results.  For simplicity, we assume spatial isotropy in this paragraph. Suppose that the hydrodynamic degrees of freedom include $N_{\mathrm{s}}$ sound modes and $N_{\mathrm{d}}$ diffusive modes.
$N_{\mathrm{d}} > 0$ generically arises when there are more even than odd conserved densities. Such imbalance densities are not `eaten' by sound modes. As we derived in Appendix \ref{app:MM}, $\mathcal{A}(\vec k)$ is proportional to an integral over the spectral weight of the charge density operator.  Because charge is always a conserved quantity, in the hydrodynamic limit we may write \cite{Kadanoff1963419} \begin{equation}
    G^{\mathrm{R}}_{\rho\rho}(\omega,k) \approx \sum_{j=1}^{N_{\mathrm{s}}} \frac{\chi_{\rho\rho}^{j\mathrm{s}}k^2}{v_{j}^2 k^2-\omega^2 - \mathrm{i}\Gamma_j\omega k^2} + \sum_{i=1}^{N_{\mathrm{d}}} \frac{D_i \chi^{i\mathrm{d}}_{\rho\rho}k^2 }{D_i k^2 - \mathrm{i}\omega} + \text{regular} \,,
\end{equation} 
where the coefficients $\chi_{\rho\rho}^{j\mathrm{s}}$ and $\chi_{\rho\rho}^{i\mathrm{d}}$ are thermodynamic susceptibilities.  Their precise form is unimportant, and for the viscous and imbalance modes that we described in the previous section, all of these susceptibilities are independent of $\ell_{\mathrm{ee}}$ and temperature. On general grounds $\Gamma_j \sim D_i \sim \ell_\text{ee}$. We then find that as $k\ell_{\mathrm{ee}} \rightarrow 0$: \begin{equation}
    \mathcal{A}(k) \approx \sum_{j=1}^{N_{\mathrm{s}}} \frac{\chi_{\rho\rho}^{j\mathrm{s}}\Gamma_j}{v_{j}^4} + \frac{1}{k^2} \sum_{i=1}^{N_{\mathrm{d}}} \frac{\chi^{i\mathrm{d}}_{\rho\rho} }{D_i} \,. \label{eq:Ak44hydro}
\end{equation}
If there is any diffusive mode, such as an imbalance mode (Section \ref{sec:2FS}), that has overlap with the density operator, so that $\chi^{i\mathrm{d}}_{\rho\rho} \neq 0$, then we immediately find that at the longest wavelength, this diffusive mode inhibits transport (makes $ \mathcal{A}(k)$ large) and so the resistivity will always increase with increasing scattering rate.   If there are no diffusive modes that couple to the charge density, as in Section \ref{sec:1FS}, then interactions always enhance transport and decrease the resistivity.

Strictly speaking, if any diffusive modes are present, the contributions to $\mathcal{A}(k)$ arising from sound modes are subleading and cannot be included, because there are $\mathcal{O}(k^2\ell_{\mathrm{ee}}^2)$ corrections to the diffusive part of the Green's function that must be accounted for.  But schematically, the hydrodynamic formula (\ref{eq:Ak44hydro}) is what our more sophisticated kinetic formalism reduces to in the limit when all hydrodynamic modes are sound or diffusion.  And of course, our kinetic formalism is also valid for models which are not isotropic or in which the hydrodynamic degrees of freedom are more complicated than simple sound waves and diffusion; the latter possibility was considered in Section \ref{sec:IN}.
}

We end with a word of caution.   Our discussion so far has focused only on the $\ell_{\mathrm{ee}}$ dependence (which could be tuned, for example, by modifying the Coulomb interaction strength via a `dielectric') and $\xi$.   In some relevant cases, the temperature dependence of the coefficients that we have set to 1 can be extremely important, because both $\ell_{\mathrm{ee}}$ and the constant prefactors that we have neglected will depend on temperature in a non-trivial way.   An important example of this is a (canonical) Galilean-invariant Fermi liquid with charge, energy and momentum conserved (in the absence of disorder).  In such a theory, $\ell_{\mathrm{ee}}\sim T^{-2}$.  At temperatures $T \ll E_{\mathrm{F}}$, one finds in the weak disorder limit, when $\xi \gg \ell_{\mathrm{ee}}$ \cite{KS11} \begin{equation}
\rho \sim \frac{\eta}{\xi^2} + \frac{Ts^2}{\kappa_{\textsc{q}}}  \sim \frac{\ell_{\mathrm{ee}}}{\xi^2} + \frac{T^2}{\ell_{\mathrm{ee}}}.
\end{equation}
We have suppressed dimensionful but temperature-independent quantities in the above formula.  This qualitative scaling can be straightforwardly recovered in our formalism.   The key point is that the term governed by thermal diffusion is suppressed at low temperatures when $\xi \sim \ell_{\mathrm{ee}}$.  Hence one will find $\rho(T)$ to be a non-monotonic function of temperature across the ballistic-to-hydrodynamic transition. \red{This is why imbalance rather than thermal modes are necessary to address the experimental challenges outlined in Section \ref{sec:intro1}.}

\section{Variational Principle for the Resistivity}
\label{sec:bounds}
In this section, we will develop a variational method suitable for upper bounding the resistivity of an inhomogeneous fluid, even when the disorder is not perturbatively weak. We begin by reviewing the technique for homogeneous fluids, and then discuss how to generalize it to inhomogeneous fluids.

\subsection{Homogeneous Fluids}
We begin with the Joule heating expression (\ref{eq:JrhoJ}) in a homogeneous fluid, where we may set $\mathsf{L}=0$.  For simplicity, assume that the electric field $\vec E$ is a unit vector in the $x$ direction, and that $\rho_{ij}$ and $\sigma_{ij}$ are (in such a coordinate basis) diagonal matrices.  It is straightforward to relax this assumption.  Then we may write \begin{equation}
\rho_{xx} = \frac{\langle \bar \Phi | \mathsf{W} | \bar\Phi\rangle}{\langle \bar\Phi |\mathsf{J}_x\rangle^2} = \frac{1}{\langle \mathsf{J}_x | \mathsf{W}^{-1}|\mathsf{J}_x\rangle},  \label{eq:rhotrue}
\end{equation}
where $\mathsf{W}|\bar\Phi\rangle  = |\mathsf{J}_x\rangle$ solves the Boltzmann equation.    There is a variational principle \cite{ziman} -- \red{that built on earlier work \cite{Kohler1948,Kohler1949,Sondheimer75}} -- which states that for any vector $|\Phi\rangle$, \begin{equation}
\rho_{xx} \le  \frac{\langle \Phi| \mathsf{W}|\Phi\rangle}{\langle \Phi|\mathsf{J}_x\rangle^2},  \label{eq:var1}
\end{equation}
with equality saturated on $|\Phi\rangle = |\bar\Phi\rangle$.  As we prove in Appendix \ref{app:joule},
\begin{equation}
\langle \bar\Phi | \mathsf{W}|\bar\Phi\rangle =  T\dot{s} 
\label{eq:WTS}
\end{equation}
where $\dot{s}$ is the entropy density production caused by Joule heating.    Hence, this variational technique admits a simple physical interpretation:  transport occurs by the pathway which minimizes entropy production, with fixed sources.     

Let us prove this variational principle.  Define \begin{equation}
\mathcal{R}[\Phi] \equiv  \frac{\langle \Phi| \mathsf{W}|\Phi\rangle}{\langle \Phi|\mathsf{J}_x\rangle^2}.
\end{equation}    Let $|\bar\Phi\rangle$ be an exact solution to (\ref{eq:LAmain}), and let $|\Phi\rangle = |\bar\Phi \rangle +  |\varphi\rangle$.  Noting that $\mathcal{R}[\lambda \Phi] = \mathcal{R}[\Phi]$ for any $\lambda \ne 0$, we may freely choose $\langle \varphi | E\rangle = 0$ by a suitable rescaling.   Then \begin{align}
\mathcal{R}[\bar\Phi + \varphi] &= \frac{\langle \bar \Phi | \mathsf{W}|\bar \Phi\rangle + 2\langle \varphi|\mathsf{W}|\bar\Phi\rangle + \langle \varphi |\mathsf{W}|\varphi\rangle}{(\langle \bar\Phi|\mathsf{J}_x\rangle + \langle \varphi|\mathsf{J}_x\rangle)^2} = \frac{\langle \bar \Phi | \mathsf{W}|\bar \Phi\rangle +
\langle \varphi |\mathsf{W}|\varphi\rangle}{\langle \bar\Phi|\mathsf{J}_x\rangle ^2}  \notag \\
&\ge \frac{\langle \bar\Phi|\mathsf{W}|\bar\Phi\rangle}{\langle \bar\Phi|\mathsf{J}_x\rangle^2}= \rho_{xx}.
\end{align}
We have used positivity of $\mathsf{W}$ as well as $\mathsf{W}|\bar\Phi\rangle = |\mathsf{J}_x\rangle$ and $\langle \mathsf{J}_x| \bar\Phi\rangle = \langle \mathsf{J}_x | \mathsf{W}^{-1}|E\rangle$ in the last step.  Hence, we always overestimate the resistivity, and the bound is saturated on the true solution to the equations of motion.

\subsection{Inhomogeneous Fluids}
We are interested  in inhomogeneous Fermi liquids, where $\mathsf{L} \ne 0$.  The presence of the antisymmetric streaming terms ruins the variational approach (\ref{eq:var1}).   It is simple to see why.  From (\ref{eq:WTS}), which holds with $\mathsf{L} \neq 0$, we see that all entropy production comes from the symmetric matrix $\mathsf{W}$, and so naively one might postulate on physical grounds that (\ref{eq:var1}) holds for both the homogeneous and inhomogeneous fluid.  However, because the matrix $\mathsf{L}$ is antisymmetric, it is a simple exercise in linear algebra to prove that for any vector,  \begin{equation}
\langle \mathsf{J}_x|\mathsf{W}^{-1}|\mathsf{J}_x\rangle \ge \langle \mathsf{J}_x|(\mathsf{W}+\mathsf{L})^{-1}|\mathsf{J}_x\rangle.
\end{equation}
Because the variational principle (\ref{eq:var1}) is minimized on (\ref{eq:rhotrue}), the true resistivity, from (\ref{eq:JrhoJ}), is larger than the minimum of the variational principle (\ref{eq:var1}).  It is not a correct variational approach.

Where does the derivation go wrong when $\mathsf{L}$ is nonzero?  Writing $|\Phi\rangle = |\bar\Phi\rangle + |\varphi\rangle$, we may still assume $\langle \varphi|\mathsf{J}_x\rangle =0$, but now $(\mathsf{W}+\mathsf{L})|\bar\Phi\rangle = |E\rangle$.  Hence, similar manipulations to before yield \begin{equation}
\mathcal{R}[\bar\Phi + \varphi] = \frac{\langle \bar\Phi| \mathsf{W}|\bar\Phi\rangle + \langle\varphi|\mathsf{W}|\varphi\rangle - 2\langle \varphi | \mathsf{L} | \bar\Phi\rangle}{\langle \bar\Phi|\mathsf{J}_x\rangle^2}.   \label{eq:varL}
\end{equation}
The final term above will generally spoil any hope of a variational principle.   However, suppose that we could force our trial functions to obey $\mathsf{L}^{\mathsf{T}}|\varphi\rangle = 0$.   Then we would again find that \begin{equation}
\mathcal{R}[\bar\Phi + \varphi]|_{\mathsf{L}^{\mathsf{T}}|\varphi\rangle = 0} = \frac{\langle \bar\Phi| \mathsf{W}|\bar\Phi\rangle + \langle\varphi|\mathsf{W}|\varphi\rangle}{\langle \bar\Phi|\mathsf{J}_x\rangle^2} \ge \rho_{xx}, 
\label{eq:boundcons}
\end{equation}
with \begin{equation}
\rho_{xx} = \frac{\langle \bar\Phi| \mathsf{W}|\bar\Phi\rangle}{\langle \bar\Phi|\mathsf{J}_x\rangle^2} = \frac{\langle \bar\Phi| \mathsf{W}+\mathsf{L}|\bar\Phi\rangle}{\langle \mathsf{J}_x | (\mathsf{W}+\mathsf{L})^{-1}|\mathsf{J}_x\rangle^2} = \frac{1}{\langle \mathsf{J}_x|(\mathsf{W}+\mathsf{L})^{-1}|\mathsf{J}_x\rangle}.  \label{eq:rhoWL1}
\end{equation}

In general, it is difficult to accomplish $\mathsf{L}^{\mathsf{T}}|\varphi\rangle = 0$, because in general $\mathsf{L}^{\mathsf{T}}|\bar\Phi\rangle \ne 0$.   In order to proceed further, it is useful to separate out the dynamics of the conserved densities and currents from the remaining modes.  Let us split the vector $|\Phi\rangle$ into three components:  \begin{equation}
    |\Phi\rangle = \left(\begin{array}{c}  |\Phi_{\mathrm{odd}}\rangle \\ |\Phi_{\mathrm{even,slow}}\rangle \\ |\Phi_{\mathrm{even,fast}}\rangle \end{array}\right).  \label{eq:52decomp}
\end{equation}
This decomposition is defined by the fact that
the matrices $\mathsf{W}$ and $\mathsf{L}$ take the form \begin{equation}
\mathsf{W} = \left(\begin{array}{ccc}  \mathsf{W}_{\mathrm{o}} &\ 0 &\ 0 \\ 0 &\ 0 &\ 0 \\ 0 &\ 0 &\ \mathsf{W}_{\mathrm{e}}\end{array}\right), \;\;\;\;\; \mathsf{L} = \left(\begin{array}{ccc}  0 &\ \mathsf{L}_1 &\ \mathsf{L}_2 \\ -\mathsf{L}_1^{\mathsf{T}} &\ 0 &\ 0 \\ -\mathsf{L}_2^{\mathsf{T}} &\ 0 &\ 0\end{array}\right),
\end{equation}
with $\mathsf{W}_{\mathrm{e}}^{-1}$ non-singular. Recall the discussion around (\ref{eq:WLEO}) above on the even/odd properties of $\mathsf{W}$ and $\mathsf{L}$. We know that $|E\rangle$ is non-vanishing only in the first (odd) component.  Using block matrix inversion identities to remove the even, fast modes, we find \begin{equation}
\langle \mathsf{J}_x|(\mathsf{W}+\mathsf{L})^{-1}|\mathsf{J}_x\rangle = \left(\begin{array}{cc} \langle \mathsf{J}_x| &\ 0 \end{array}\right) \left(\begin{array}{cc} \mathsf{W}_{\mathrm{o}} + \mathsf{L}_2 \mathsf{W}_{\mathrm{e}}^{-1} \mathsf{L}_2^{\mathsf{T}} &\ \mathsf{L}_1 \\ -\mathsf{L}_1^{\mathsf{T}} &\ 0 \end{array}\right)^{-1}\left(\begin{array}{c} |\mathsf{J}_x\rangle \\ 0\end{array}\right).  \label{eq:blockWL}
\end{equation}
The matrix $\mathsf{W}_{\mathrm{o}} + \mathsf{L}_2 \mathsf{W}_{\mathrm{e}}^{-1} \mathsf{L}_2^{\mathsf{T}}$ is manifestly positive-definite and symmetric.   Now suppose that the list of even conserved quantities (null vectors of $\mathsf{W}$) is finite (at each point $x$).   Then, the middle row of $(\mathsf{W}+\mathsf{L})|\Phi\rangle = |E\rangle$ (using the three-block decomposition of (\ref{eq:52decomp})) implies that: 
\begin{equation}
\mathsf{L}_1^{\mathsf{T}}|\Phi_{\mathrm{odd}}\rangle = 0.  \label{eq:L1TPhi}
\end{equation}
Since this equation must be true on the background solution, writing $|\Phi_{\mathrm{odd}} \rangle = |\bar\Phi_{\mathrm{odd}}\rangle + |\varphi_{\mathrm{odd}}\rangle$, we conclude that imposing (\ref{eq:L1TPhi}) on a trial wave function necessarily imposes \begin{equation}
\mathsf{L}_1^{\mathsf{T}}|\varphi_{\mathrm{odd}}\rangle = 0.  \label{eq:L1Tphi}
\end{equation}
So long as there are only a finite number of conserved quantities, then we will only have a finite number of these constraint equations. 

Using the even/odd decomposition, the variation (\ref{eq:varL}) can be written
\begin{equation}
\mathcal{R} = \frac{\langle \bar\Phi_{\mathrm{odd}}| \widetilde{\mathsf{W}}|\bar\Phi_{\mathrm{odd}}\rangle + \langle\varphi_{\mathrm{odd}}|\widetilde{\mathsf{W}}|\varphi_{\mathrm{odd}}\rangle - 2\langle \varphi_{\mathrm{odd}} | \mathsf{L}_1 | \bar\Phi_{\mathrm{even,slow}}\rangle -2\langle \bar\Phi_{\mathrm{odd}} | \mathsf{L}_1 | \varphi_{\mathrm{even,slow}}\rangle}{\langle \bar\Phi|\mathsf{J}_x\rangle^2} \,,
\end{equation}
where
\begin{equation}
\widetilde{\mathsf{W}} \equiv \mathsf{W}_{\mathrm{o}} + \mathsf{L}_2 \mathsf{W}_{\mathrm{e}}^{-1} \mathsf{L}_2^{\mathsf{T}}.
\end{equation}
Assuming the constraints (\ref{eq:L1TPhi}) and (\ref{eq:L1Tphi}), we observe that both antisymmetric terms above vanish regardless of the variational choice of $|\Phi_{\mathrm{even,slow}}\rangle$.  Therefore, following the same logic as (\ref{eq:boundcons}), we are led to the following constrained variational principle for the resistivity: \begin{equation}
\rho_{xx} \le \widetilde{\mathcal{R}}[\widetilde\Phi] \equiv \left.\frac{\langle \widetilde \Phi| \widetilde{\mathsf{W}}|\widetilde \Phi\rangle}{\langle \widetilde\Phi|\mathsf{J}_x\rangle^2} \right|_{\mathsf{L}_1^{\mathsf{T}}|\widetilde\Phi\rangle = 0},\label{eq:bound32}
\end{equation}
with $\widetilde \Phi$ running only over odd vectors.  At present, this looks abstract and possibly useless.   However, we expect that in many problems of interest, the only even zero modes of $\mathsf{W}$ correspond to scalar (under spatial rotations) conserved quantities such as energy or charge, where (\ref{eq:L1TPhi}) is nothing more than $\nabla \cdot \vec{\mathcal{J}} =0$, for each conserved current $\vec{\mathcal{J}}$. 

Hence, the resistivity is then bounded from above by the rate of entropy production, subject to the constraints that all currents associated with conserved quantities are exactly divergenceless.  In the hydrodynamic limit, this is equivalent to the hydrodynamic transport bounds of \cite{Lucas:2015lna, Grozdanov:2015qia, Grozdanov:2015djs,ushydro}, which themselves are generalizations of Thomson's principle, a variational approach for computing the effective resistance of a resistor network \cite{levin}.   The advantage of the kinetic variational principle over the hydrodynamic ones is (in addition to the distinct, but overlapping, regime of applicability) the fact that we do not need to know an explicit expression for the dissipative hydrodynamic coefficients such as viscosity.  

\subsection{How Interactions Modify Transport}
We have presented a general variational principle (\ref{eq:bound32}).  To make specific quantitative predictions, more information about $\mathsf{W}$ is necessary.   Nonetheless, following the discussion of Section \ref{sec:genprin}, let us make a few general comments -- now at the non-perturbative level.

In the ballistic (non-interacting) limit where $\mathsf{W}\rightarrow 0$, it is challenging to directly deduce from (\ref{eq:bound32}) that the resistivity saturates to a constant \begin{equation}
\rho_{xx} \lesssim \frac{C}{\tilde \nu \tilde v_{\mathrm{F}}\xi},  \label{eq:53xi}
\end{equation}
with $\tilde \nu$ and $\tilde v_{\mathrm{F}}$ the `averaged' density of states and velocity of quasiparticles, respectively, and $C$ a dimensionless number.  See Appendix \ref{app:varbal} for details of a direct variational calculation which confirms this.   
Assuming that the resistivity is finite, it is simple to show the scaling (\ref{eq:53xi}).  When $\mathsf{W} = 0$, if we define the coordinate $X = x/\xi$, then the Boltzmann equation becomes \begin{equation}
\mathsf{L}|\Phi\rangle =  \frac{1}{\xi} \left|\vec v\cdot \frac{\partial \Phi}{\partial \vec X} - \frac{\partial V_{\mathrm{imp}}}{\partial \vec X} \cdot \frac{\partial \Phi}{\partial \vec p}\right\rangle = 0.
\end{equation}     
Because $\mathsf{L}$ exactly scales as $1/\xi$, we deduce that $(z+\mathsf{L})^{-1} \sim \xi$.   Hence from the expression (\ref{eq:rhoWL1}) for the resistivity, we conclude that (\ref{eq:53xi}) must follow on dimensional grounds alone. 

In the limit when interactions become important, it is clear from the form of $\widetilde{\mathsf{W}}$ that our goal is to find an admissible $|\widetilde \Phi\rangle$ for the variational principle which is -- as much as possible -- exactly conserved in the absence of disorder:  $\mathsf{W}_{\mathrm{o}}|\widetilde \Phi\rangle = 0$.  If we can achieve this, while remaining consistent with all current conservation laws, then the resistivity \begin{equation}
\rho_{xx} \lesssim \frac{\langle \widetilde \Phi| \mathsf{L}_2 \mathsf{W}_{\mathrm{e}}^{-1} \mathsf{L}_2^{\mathsf{T}}|\widetilde \Phi\rangle }{\langle \widetilde \Phi|\mathsf{J}_x\rangle^2} \sim \frac{\ell_{\mathrm{ee}}}{\xi^2}.
\end{equation}
We have used the exact scaling $\mathsf{L} \sim 1/\xi$, as well as the heuristic scaling $\mathsf{W}_\text{e} \sim 1/\ell_{\mathrm{ee}}$.   Clearly, the resistivity decreases as the interaction strength increases (and $\ell_{\mathrm{ee}}$ decreases).

However, we expect that in general, there will be more conserved currents than odd conserved quantities.  This is the case in a conventional fluid, where there is both charge and energy conservation, but only momentum conservation.  As we noted previously, of additional importance in many solid-state systems are imbalance modes.  If it is not possible to satisfy all conservation laws on an ansatz $|\widetilde \Phi\rangle$ with $\mathsf{W}_{\mathrm{o}}|\widetilde \Phi\rangle = 0$, then we expect that in general the resistivity is dominated by the microscopic scattering rates: \begin{equation}
\rho \lesssim \frac{\langle \widetilde \Phi|  \mathsf{W}_{\mathrm{o}} |\widetilde \Phi\rangle }{\langle \widetilde \Phi|\mathsf{J}_x\rangle^2} \sim \frac{1}{\ell_{\mathrm{ee}}}.
\end{equation}
In this limit, interactions enhance the resistivity and the scattering rate for momentum will appear to be set by $\ell_{\mathrm{ee}}$, although (as we have explained) this is not quite the correct physical interpretation.

This discussion gives us non-perturbative confidence that the bounds described in Section \ref{sec:summary} are qualitatively correct.   There are some special limiting cases where in hydrodynamic limits \cite{KS11},  one may look for ansatzes $|\Phi\rangle$ where $\rho$ increases more slowly than $\ell_{\mathrm{ee}}^{-1}$ in the limit $\ell_{\mathrm{ee}} \ll \xi$.   These ansatzes are similar to those discussed in Appendix \ref{app:varbal}.  Based on the discussion in Section \ref{sec:genprin}, such a limit must be non-perturbative in the disorder strength.   If there are at least $d$ even conserved modes which must diffuse, and momentum is the only inversion-odd conserved quantity, we do not expect that such special ansatzes will be able to parametrically reduce the resistivity of a metal in $d$ spatial dimensions.   


\section{Conclusion}
We have described how to solve the inhomogeneous Boltzmann equation to compute the electrical resistivity of a metal.  In appropriate limits, we  have reproduced ballistic and (conventional, \red{viscous}) hydrodynamic transport.  We have also discovered a variety of novel effects, which we can often interpret with a generalized hydrodynamics \cite{ushydro}. \red{At a qualitative level, our main observation is that if there are more inversion-even conserved quantities than inversion-odd conserved quantities, then textbook viscous hydrodynamics does not emerge in the limit that the mean free path is short compared to the disorder wavelength. Instead the dynamics is dominated by diffusive `imbalance' (or other such) modes. In such regimes the interactions cause the resistivity to increase, in dramatic contrast to the viscous case.} When quasiparticles are well-defined, we \red{have provided two practical methods for solving  the full kinetic transport problem in condensed matter physics: the perturbative framework of Section \ref{sec:perturb} and the variational methods of Section \ref{sec:bounds}.}


 Momentum can be relaxed by short-range quantum impurity scattering and umklapp processes.  These can be accounted for by adding terms to $\mathsf{W}$ which do not conserve momentum, in the conventional way.   The formal perturbative expansion in Section \ref{sec:perturb} is broken by such collisions.  Nonetheless, the variational formalism given in Section \ref{sec:bounds} remains valid whether or not collisions relax momentum, and it suggests that the existence of a small amount of umklapp or quantum impurity scattering will only weakly modify the results described above. If $\ell_{\mathrm{um}}$ is the mean free path for umklapp processes, for example, there will be an intermediate `hydrodynamic' regime with imbalance diffusion when $\ell_{\mathrm{ee}} \ll \xi \ll \ell_{\mathrm{um}}$.  If umklapp can relax imbalance modes, then a conventional hydrodynamic regime will only emerge when $\xi \gg \ell_{\mathrm{um}}$, similar to what we saw in Section \ref{sec:IN}.

In the future, it will be important to extend this model to magnetotransport and nonzero frequency transport, as well as model other types of disorder. We look forward to extensions of this formalism, along with applications to specific materials and to experimental tests of our predictions.

\addcontentsline{toc}{section}{Acknowledgements}
\section*{Acknowledgements}
We thank Caleb Cook, and especially Steven Kivelson and  Leonid Levitov, for helpful discussions.  AL was supported by the Gordon and Betty Moore Foundation's EPiQS Initiative through Grant GBMF4302.  SAH is partially supported by a DOE Early Career Award.

\begin{appendix}

\section{Memory Matrix Formalism}
\label{app:MM}

We are studying a quantum many-body system deformed by a perturbatively small chemical potential.  The microscopic Hamiltonian is given by  \begin{equation}
H = H_{\mathrm{clean}} + \int \mathrm{d}^dx \; n(\vec x) V_{\mathrm{imp}}(\vec x),
\end{equation}
with $n(\vec x)$ now interpreted as the charge density operator.   It is now very well understood that if momentum $P$ is the only inversion-odd conserved quantity of $H_{\mathrm{clean}}$, and $V_{\mathrm{imp}}$ is perturbatively small, that the resistivity is given by \cite{Hartnoll:2016apf, Hartnoll:2007ih, Hartnoll:2012rj, Lucas:2015pxa, Lucas:2015lna}: \begin{equation}
\rho_{ij} = \frac{1}{\chi_{JP}^2} \lim_{\omega\rightarrow 0} \frac{\mathrm{Im}\left(G^{\mathrm{R}}_{\dot{P}_i \dot{P}_j}(k,\omega)\right)}{\omega} = \frac{1}{\chi_{JP}^2} \int \frac{\mathrm{d}^dk }{(2\pi)^d} k_i k_j |V_{\mathrm{imp}}(k)|^2 \lim_{\omega\rightarrow 0} \frac{\mathrm{Im}\left(G^{\mathrm{R}}_{nn}(k,\omega)\right)}{\omega}.  \label{eq:MM1}
\end{equation}
$\chi_{JP}$ is the susceptibility between the charge current and the momentum, and we can evaluate the Green's function $G^{\mathrm{R}}$ in the clean theory.  One can think of this as a generalization of Fermi's golden rule to interacting quantum systems.

We first write the operator \begin{equation}
n = \sum_{\vec p} n_{\vec p}.
\end{equation}
In the kinetic limit where $k \ll k_{\mathrm{F}}$, we may approximate retarded Green's functions using kinetic theory, using the general technique of \cite{Kadanoff1963419}:  if the equations of motion for `long-lived' quantities $\varphi_a$ take the form \begin{equation}
    \partial_t \delta \varphi_a + M_{ab}\delta \varphi_b = 0,
\end{equation} 
and the susceptibility matrix of the $\varphi_a$ is $\chi_{ab}$, then at low frequencies \begin{equation}
    G^{\mathrm{R}}_{ad}(\vec k, \omega) = M_{ab}(\vec k) \left[M(\vec k) - \mathrm{i}\omega\right]^{-1}_{bc}\chi_{cd}\,,
\end{equation}
where $bc$ refers to the components of the inverted matrix.
The spectral weight is then \begin{equation}
    \lim_{\omega\rightarrow 0} \frac{\mathrm{Im}\left(G^{\mathrm{R}}_{ab}(\vec k,\omega)\right)}{\omega} = M^{-1}_{ac}(\vec k) \chi_{cb}.
\end{equation}
We take the $\varphi_a$ to be the number density of quasiparticles at momentum $\vec p$. In the clean theory, the equations of motion read \begin{equation}
    \partial_t \delta n_{\vec p} +\left(-\frac{\partial f_{\mathrm{eq}}}{\partial \epsilon}\right) (\mathsf{W}+\mathsf{L})\left(-\frac{\partial f_{\mathrm{eq}}}{\partial \epsilon}\right)^{-1}\delta n_{\vec p}=0,
\end{equation}
(recall that $\delta n_{\vec p} = (-\partial_\epsilon f_{\mathrm{eq}})_{\vec p} \Phi_{\vec p}$) which gives \begin{equation}
    M_{n_{\vec p} n_{\vec q}} = \left(-\frac{\partial f_{\mathrm{eq}}}{\partial \epsilon}\right)_{\vec p}(\mathsf{W}+\mathsf{L})_{\vec p \vec q}\left(-\frac{\partial f_{\mathrm{eq}}}{\partial \epsilon}\right)^{-1}_{\vec q}  \label{eq:MMM}
\end{equation}
The susceptibility we require is \begin{equation}
\chi_{n,n_{\vec p}} = \frac{\partial n_{\vec p}}{ \partial \mu} = \left(-\frac{\partial f_{\mathrm{eq}}}{\partial \epsilon}\right)_{\vec p}.  \label{eq:MMX}
\end{equation}The first step in this equation follows from the fact that $\mu$ is the thermodynamic potential conjugate to $n$ \cite{Kadanoff1963419}.   Combining (\ref{eq:MM1}), (\ref{eq:MMM}) and (\ref{eq:MMX}) with our non-trivial inner product (\ref{eq:innerproduct}), we obtain agreement with (\ref{eq:sigmaexact4}) in the main text.

The generalization of (\ref{eq:sigmaexact4}) to the case where there are multiple conserved inversion-odd quantities is immediate within the memory matrix formalism \cite{Hartnoll:2016apf}.  Let us denote with $Q_a$ the list of odd conserved quantities, of which a finite subset are the momenta $P_i$.   Then  the conductivity tensor $\sigma_{ij}$ is given by \begin{equation}
\sigma_{ij} = \chi_{J_i Q_a} \left(\lim_{\omega\rightarrow 0} \frac{\mathrm{Im}\left(G^{\mathrm{R}}_{\dot{Q}_a \dot{Q}_b}(k,\omega)\right)}{\omega}\right)^{-1}   \chi_{J_j Q_b}   \label{eq:multcons}
\end{equation}
where the spectral weight is to be inverted as a matrix with $ab$ indices.   To obtain $\rho_{ij}$, one takes the inverse of $\sigma_{ij}$ over $ij$ indices as usual.   If the disorder does not break every conservation law, and so there is some $\dot{Q}_a=0$,  then the conductivity will be infinite.

\section{Toy Models of 2d Fermi Liquids at Weak Disorder}
\label{app:toy1}
The toy models that we have presented are exactly solvable due to an elegant mathematical trick introduced in \cite{levitov2}, and extended in \cite{2016arXiv161200856L, levitov3}.   Let us write (for the homogeneous theory) \begin{equation}
    \mathsf{W} = \mathsf{W}_0 - \mathsf{X},
\end{equation}
where $\mathsf{W}_0$ is chosen such that $(\mathsf{W}_0+\mathsf{L}(\vec k))^{-1} \equiv \mathsf{G}(\vec k)$ is analytically computable; we will often leave the $\vec k$ dependence implicit.  Suppose that $\mathsf{P}$ projects onto a finite number of modes, and that $\mathsf{X} = \mathsf{PXP}$.    Then we write \begin{align}
(\mathsf{W}+\mathsf{L})^{-1} &= (\mathsf{W}_0+\mathsf{L}-\mathsf{X})^{-1} = \sum_{n=0}^\infty \mathsf{G}\left(\mathsf{PXPG}\right)^n = \mathsf{G} + \mathsf{GPX}\left(\sum_{n=0}^\infty (\mathsf{PGPX})^n \right)\mathsf{PG}  
\end{align}
Defining $\widetilde{\mathsf{G}} \equiv \mathsf{PGP}$, we find \begin{equation}
(\mathsf{W}+\mathsf{L})^{-1} = \mathsf{G} + \mathsf{GP}\left[\mathsf{X}\left(1-\widetilde{\mathsf{G}}\mathsf{X}\right)^{-1}
\right]\mathsf{PG}.  \label{eq:WLtrick}
\end{equation}
The only matrix inverse we must compute, which is located within the square brackets above, is of a finite dimensional matrix.   Hence, it is highly efficient to compute $(\mathsf{W}+\mathsf{L})^{-1}$ numerically.  In fact, in simple cases, we can compute it analytically.

Let us begin with the simple model of a single rotationally symmetric Fermi surface for a 2d Fermi liquid at low temperature, discussed in Section \ref{sec:1FS}.   This will form the basis for all the toy models we consider in this paper.  In this model, \begin{equation}
   \mathsf{W}_0 = \frac{1}{\nu(\mu)}\times \frac{v_{\mathrm{F}}}{\ell_{\mathrm{ee}}}, \;\;\;\;  \mathsf{X} = \frac{1}{\nu(\mu)}\times\frac{v_{\mathrm{F}}}{\ell_{\mathrm{ee}}} \mathsf{P}, \;\;\;\; \mathsf{P} = \frac{1}{\nu(\mu)}\times\left(|1\rangle\langle 1| + |0\rangle\langle 0| + |-1\rangle\langle -1|\right).
\end{equation}
The overall prefactors of $1/\nu$ follow from (\ref{eq:toymodelinner}).  Similarly, one can compute \begin{equation}
    \mathsf{L}(\vec k) = \sum_n\frac{v_{\mathrm{F}}}{2\nu(\mu)} \times \left((\mathrm{i}k_x +k_y)|n-1\rangle\langle n| + (\mathrm{i}k_x - k_y)|n+1\rangle\langle n| \right).
\end{equation}
Using contour integration one can exactly compute \cite{2016arXiv161200856L} \begin{equation}
\mathsf{G}_{mn}(\vec k) \equiv G^0_{mn}(\vec k) =\frac{1}{\nu(\mu)}\times  \frac{(-\mathrm{i})^{|n-m|}\mathrm{e}^{\mathrm{i}(n-m)\phi_k}}{\sqrt{k^2+\ell_{\mathrm{ee}}^{-2}}}\left(\sqrt{1+\frac{1}{(k\ell_{\mathrm{ee}})^2}} - \frac{1}{k\ell_{\mathrm{ee}}}\right)^{|n-m|},  \label{eq:Gmn}
\end{equation}
with $|k| \mathrm{e}^{\mathrm{i}\phi_k} \equiv k_x + \mathrm{i}k_y$.  As noted in the main text, we normalize the harmonic bras $\langle j|j^\prime\rangle = \mdelta_{jj^\prime}$; hence, the non-trivial inner product is not relevant for the matrix multiplication in (\ref{eq:WLtrick}).  It will become relevant for computing (\ref{eq:sigmaexact4}), because there we did not normalize the bras.   In this simple model of a Fermi surface, the unnormalized and normalized inner products differ only by a factor $\nu$.   Given $\widetilde{\mathsf{G}}$ and $\mathsf{X}$, we may now  compute $\mathcal{A}(k) = \nu \langle 0| (\mathsf{W}+\mathsf{L})^{-1}|0\rangle$;  the result is shown in (\ref{eq:432}) in the main text.   

It is simple to generalize to the more complicated models.   In the  model of Section \ref{sec:IN},  one uses an identical $\mathsf{G}$ as before, but sets \begin{equation}
    \mathsf{W}_0 =\frac{1}{\nu(\mu)}\times  \frac{\nu v_{\mathrm{F}}}{\ell_{\mathrm{ee}}}, \;\;\;\; \mathsf{X} =\frac{1}{\nu(\mu)}\times  \frac{\nu v_{\mathrm{F}}}{\ell_{\mathrm{ee}}} \left(\mathsf{P}-b|2\rangle\langle 2| - b|-2\rangle\langle -2|\right), \;\;\;\; \mathsf{P} =\frac{1}{\nu(\mu)}\times \sum_{|j|\le 2} |j\rangle\langle j|.
\end{equation}
Again, the finite dimensional matrix inverse in (\ref{eq:WLtrick}) can be done analytically, and the result for $\mathcal{A}(k)\sim \langle 0| (\mathsf{W}+\mathsf{L})^{-1}|0\rangle$ is shown in (\ref{eq:433}) in the main text.    In the model of two Fermi surfaces (Section \ref{sec:2FS}) and of electron-phonon coupling (Section \ref{sec:EPH}), by rewriting (\ref{eq:L434}) as \begin{equation}
\mathsf{L} =\frac{1}{\nu(\mu)}\times  \mathrm{i}v_{\mathrm{F,2}}(\cos\theta k_x + \sin \theta k_y) \left(\mathsf{P}_2 + \zeta \mathsf{P}_1\right) \,,
\end{equation} 
with \begin{equation}
\zeta \equiv \frac{v_{\mathrm{F,1}}}{v_{\mathrm{F,2}}},
\end{equation}
it is straightforward to see that, upon choosing $\mathsf{W}_0 =\nu v_{\mathrm{F,2}}/\ell_{\mathrm{ee}}$ as before,   \begin{equation}
\mathsf{G} = \sum_{mn} \left(G^0_{mn}(\zeta \vec k) |m1\rangle \langle n1| + G^0_{mn}(\vec k) |m2\rangle\langle n2|\right).
\end{equation}
Now, defining \begin{equation}
\cos\alpha \equiv \frac{1}{\sqrt{1+\zeta^2}},
\end{equation}
we take \begin{equation}
\mathsf{X} =\frac{1}{\nu(\mu)}\times  \frac{ v_{\mathrm{F,2}}}{\ell_{\mathrm{ee}}} \left[\hat b|01\rangle\langle 01| + |02\rangle\langle 02| +  \sum_{j=\pm 1} (\sin\alpha |j1\rangle + \cos\alpha |j2\rangle)(\sin\alpha \langle j1| + \cos\alpha \langle j2|) \right].
\end{equation}where $\hat b = 0$ in the electron-phonon model, and $\hat b=1$ in the model of two Fermi surfaces.  $\mathsf{X}$ hence is a projector onto the conserved quantities, in each model.  It is straightforward to numerically compute $\mathcal{A}(\vec k)$ from here.

\section{Consequences of Inversion Symmetry at Weak Disorder}
\label{app:genprin}
We explicitly carry out the matrix inverse for $\mathsf{W}+\mathsf{L}$ given in  (\ref{eq:4x4}).  First integrating out only the fast modes (but not the odd slow modes) we obtain 
\begin{subequations}\label{eq:4x4C}\begin{align}
\widehat{\mathsf{W}}_{\mathrm{e}} &= \mathsf{L}^{\mathsf{T}}_{\mathrm{of,es}}(\mathsf{W}_{\mathrm{o}} + \mathsf{L}_{\mathrm{ef,of}}^{\mathsf{T}} \mathsf{W}_{\mathrm{e}}^{-1} \mathsf{L}_{\mathrm{ef,of}})^{-1}\mathsf{L}_{\mathrm{of,es}}, \\
\widehat{\mathsf{W}}_{\mathrm{o}} &= \mathsf{L}^{\mathsf{T}}_{\mathrm{ef,os}}(\mathsf{W}_{\mathrm{e}} + \mathsf{L}_{\mathrm{of,ef}}^{\mathsf{T}} \mathsf{W}_{\mathrm{o}}^{-1} \mathsf{L}_{\mathrm{of,ef}})^{-1}\mathsf{L}_{\mathrm{ef,os}}, \\
\widehat{\mathsf{L}}_{\mathrm{os,es}} &=\mathsf{L}_{\mathrm{os,es}} - \mathsf{L}^{\mathsf{T}}_{\mathrm{ef,os}} (\mathsf{W}_{\mathrm{e}} + \mathsf{L}_{\mathrm{of,ef}}^{\mathsf{T}} \mathsf{W}_{\mathrm{o}}^{-1} \mathsf{L}_{\mathrm{of,ef}})^{-1}\mathsf{L}_{\mathrm{ef,of}}\mathsf{W}^{-1}_{\mathrm{o}}\mathsf{L}_{\mathrm{of,es}} \,, \\
\widehat{\mathsf{L}}_{\mathrm{es,os}} &= \mathsf{L}_{\mathrm{es,os}} - \mathsf{L}^{\mathsf{T}}_{\mathrm{of,es}} (\mathsf{W}_{\mathrm{o}} + \mathsf{L}_{\mathrm{ef,of}}^{\mathsf{T}} \mathsf{W}_{\mathrm{e}}^{-1} \mathsf{L}_{\mathrm{ef,of}})^{-1}\mathsf{L}_{\mathrm{of,ef}}\mathsf{W}^{-1}_{\mathrm{e}}\mathsf{L}_{\mathrm{ef,os}} = -\widehat{\mathsf{L}}_{\mathrm{os,es}}^{\mathsf{T}}.
\end{align}\end{subequations}

Next, we assume for simplicity that  $\widehat{\mathsf{W}}_{\mathrm{o}}$ is invertible.   We find (\ref{eq:Ak44}) as in the main text, and so all that remains is to justify (\ref{eq:Weoscaling})  and (\ref{eq:Leoscaling}).  In order to derive these scaling arguments, let us explicitly define $\ell_{\mathrm{ee}}$ and $v_{\mathrm{F}}$ as follows: \begin{subequations}\label{eq:appcscale}
\begin{align}
    kv_{\mathrm{F}} &\sim  \text{eigenvalue of } \mathsf{L} = \mathrm{i}\vec k \cdot \vec v, \\
    \frac{v_{\mathrm{F}}}{\ell_{\mathrm{ee}}} &\sim \text{typical eigenvalue of } \mathsf{W}_{\mathrm{e,o}}.
\end{align}
\end{subequations}
In both of the equations above, and for the remainder of the appendix, we have used $\sim$ to denote that we are neglecting O(1) prefactors.   Both eigenvalues should be computed in the absence of disorder;  we have already integrated out disorder by (\ref{eq:sigmaexact4}).   In models with parametric hierarchies of relaxation times, or both very fast and very slow fermions, the scalings above should not be expected to fully capture the physics at intermediate length scales, and one would need to define a larger block matrix decomposition than (\ref{eq:4x4}) to keep track of modes that decay at parametrically different rates, for example.

First, let us discuss the hydrodynamic limit.  In this case, because by definition both $\mathsf{W}_{\mathrm{e}}$ and $\mathsf{W}_{\mathrm{o}}$ are invertible, we may approximate   \begin{equation}
      \widehat{\mathsf{W}}_{\mathrm{e}} \approx \mathsf{L}^{\mathsf{T}}_{\mathrm{of,es}}\mathsf{W}_{\mathrm{o}}^{-1}\mathsf{L}_{\mathrm{of,es}} + \mathcal{O}\left(k^2\ell_{\mathrm{ee}}^2\right)  \sim k^2 \ell_{\mathrm{ee}} v_{\mathrm{F}}.
\end{equation}
A similar equation holds for $\widehat{\mathsf{W}}_{\mathrm{o}}$.   We next find that 
\begin{equation}
\widehat{\mathsf{L}}_{\mathrm{os,es}} \approx \mathsf{L}_{\mathrm{os,es}} - \mathsf{L}^{\mathsf{T}}_{\mathrm{ef,os}} \mathsf{W}_{\mathrm{e}}^{-1}\mathsf{L}_{\mathrm{ef,of}}\mathsf{W}^{-1}_{\mathrm{o}}\mathsf{L}_{\mathrm{of,es}} + \mathcal{O}\left(k^2\ell_{\mathrm{ee}}^2\right) \sim k v_{\mathrm{F}} \left(1+ k^2 \ell_{\mathrm{ee}}^2\right) \sim kv_{\mathrm{F}}. 
\end{equation}
A similar equation holds for $\widehat{\mathsf{L}}_{\mathrm{es,os}}$.

The ballistic limit is somewhat more subtle:  the infinite dimensionality of the vector space of fast modes is critical to obtain physically sensible results.   Consider the matrix $\mathsf{W}_{\mathrm{o}} + \mathsf{L}_{\mathrm{ef,of}}^{\mathsf{T}} \mathsf{W}_{\mathrm{e}}^{-1} \mathsf{L}_{\mathrm{ef,of}}$, which must be inverted to compute $\widehat{\mathsf{W}}_{\mathrm{e}}$.   A naive application of (\ref{eq:appcscale}) suggests that this matrix scales as $\ell_{\mathrm{ee}}k^2$ and hence that $\widehat{\mathsf{W}}_{\mathrm{e}} \sim \ell_{\mathrm{ee}}^{-1}$ -- namely, that in the ballistic limit there is no dissipation at all.    However, this is not true.  At a mathematical level, the argument above fails because the matrix $\mathsf{L}_{\mathrm{of,ef}}$ is very far from full rank, and so has many null vectors.  Physically, these null vectors correspond to any velocity for which $\vec k \cdot \vec v = 0$.  Do these null vectors lead to $\widehat{\mathsf{W}}_{\mathrm{e}} \sim \ell_{\mathrm{ee}}k^2$, which would diverge in the ballistic limit?  This is also not correct, because although there are infinitely many null vectors, they still form a set of measure zero of the total ``size" of the vector space.  Following the explicit calculation of Section \ref{sec:free}, we expect that the vanishing eigenvalues of a vanishingly small fraction of eigenvectors lead instead to
\begin{equation}
    \widehat{\mathsf{W}}_{\mathrm{e}} \sim kv_{\mathrm{F}} + \mathcal{O}\left(\frac{1}{\ell_{\mathrm{ee}}}\right).  \label{eq:appCbal}
\end{equation}
This result can alternatively be understood by observing that $(\mathsf{W}+\mathsf{L})^{-1}$ is not a singular matrix in the ballistic limit,\footnote{This is easiest to see in position space, where one can see by directly solving the Liouville (non-interacting Boltzmann) equation that $\langle \vec x \vec p | (\mathsf{W}+\mathsf{L})^{-1}|\vec x_0 \vec p_0\rangle \propto \mathrm{\Theta}(\vec v(\vec p) \cdot (\vec x - \vec x_0)) \mdelta(\vec p-\vec p_0) \mdelta(\vec v\times (\vec x - \vec x_0))$.}  and so by dimensional analysis one is forced to arrive at (\ref{eq:appCbal}).  There is less  subtlety in estimating $\widehat{\mathsf{L}}_{\mathrm{os,es}}$ in the ballistic limit.   A naive application of (\ref{eq:appcscale}), together with (\ref{eq:4x4C}), leads to \begin{equation}
    \widehat{\mathsf{L}}_{\mathrm{os,es}} \sim kv_{\mathrm{F}}.
\end{equation}
As we saw above, this is more precisely argued for by noting that all components of $(\mathsf{W}+\mathsf{L})^{-1}_{\text{fast}} \sim 1/kv_{\mathrm{F}}$.  The end result is the same.

Combining the results of the previous two paragraphs, we arrive at (\ref{eq:Weoscaling}) and (\ref{eq:Leoscaling}).

$\widehat{\mathsf{W}}_{\mathrm{o}}$ is invertible if the streaming terms couple every odd mode to an even mode.   This is not always the case -- see for example the model of a conserved $j=2$ harmonic in Section \ref{sec:IN}.  If this assumption is not satisfied, then we can modify the block decomposition of (\ref{eq:4x4}) straightforwardly to \begin{equation}
|\Phi\rangle = \left(\begin{array}{c} |\Phi_{\mathrm{even,slow}}\rangle \\ |\Phi_{\mathrm{odd,slow}^\prime}\rangle \\ |\Phi_{\mathrm{odd,slow}}\rangle  \\  |\Phi_{\mathrm{even,fast}}\rangle \\ |\Phi_{\mathrm{odd,fast}}\rangle \end{array}\right), \;\;\;\;\; \mathsf{W}+\mathsf{L} = \left(\begin{array}{ccccc} 0 &\ \mathsf{L}_{\mathrm{s1}} &\  \mathsf{L}_{\mathrm{es,os}} &\ 0 &\ \mathsf{L}_{\mathrm{es,of}} \\ -\mathsf{L}_{\mathrm{s1}}^{\mathsf{T}} &\ 0 &\ 0 &\ 0 &\ 0 \\ \mathsf{L}_{\mathrm{os,es}} &\ 0 &\ 0 &\ \mathsf{L}_{\mathrm{os,ef}} &\ 0 \\ 0 &\ 0 &\ \mathsf{L}_{\mathrm{ef,os}} &\ \mathsf{W}_{\mathrm{e}} &\ \mathsf{L}_{\mathrm{ef,of}} \\ \mathsf{L}_{\mathrm{of,es}} &\ 0 &\ 0 &\ \mathsf{L}_{\mathrm{of,ef}} &\ \mathsf{W}_{\mathrm{o}}  \end{array}\right) \,. \label{eq:5x5}
\end{equation}
We have added a fifth row and column to this block matrix that corresponds to the odd conserved quantities $|\Phi_{\mathrm{odd,slow}^\prime}\rangle$ which do not couple via streaming terms to fast modes.  We can again integrate out the fast degrees of freedom, which leads to the following modification of (\ref{eq:44top22}): \begin{equation}
    (\mathsf{W}+\mathsf{L})^{-1}_{\mathrm{slow}} = \left(\begin{array}{ccc} \widehat{\mathsf{W}}_{\mathrm{e}} &\ \mathsf{L}_{\mathrm{s1}} &\ \widehat{\mathsf{L}}_{\mathrm{es,os}} \\ -\mathsf{L}_{\mathrm{s1}}^{\mathsf{T}} &\ 0 &\ 0 \\ \widehat{\mathsf{L}}_{\mathrm{os,es}} &\ 0 &\ \widehat{\mathsf{W}}_{\mathrm{o}} \end{array}\right)^{-1} \,,
\end{equation}  together with (\ref{eq:4x4C}).   The generalization of (\ref{eq:Ak44}) is \begin{equation}
    \mathcal{A})(\vec k) = \langle \mathsf{n} | \mathsf{M} - \mathsf{M} \mathsf{L}_{\mathrm{s1}} \left(\mathsf{L}_{\mathrm{s1}}^{\mathsf{T}}\mathsf{M}\mathsf{L}_{\mathrm{s1}}\right)^{-1} \mathsf{L}_{\mathrm{s1}}^{\mathsf{T}}\mathsf{M} |\mathsf{n}\rangle,  \text{ with } \mathsf{M} = \left(\widehat{\mathsf{W}}_{\mathrm{e}} + \widehat{\mathsf{L}}_{\mathrm{es,os}}\widehat{\mathsf{W}}^{-1}_{\mathrm{o}} \widehat{\mathsf{L}}_{\mathrm{es,os}}^{\mathsf{T}}\right)^{-1}.
\end{equation} 
The presence of the additional odd conservation laws thus reduces the spectral weight.   In particular, the role of $\mathsf{L}_{\mathrm{s1}}$ in the above formula is, roughly, to project out the components of $\mathsf{M}$. which have overlap with the odd slow modes.   This may be enough to kill any imbalance modes in $\mathsf{M}$ (eigenvalues $\sim k^{-2}$); a more detailed analysis requires more specific details of the model.

The case we studied in Section \ref{sec:IN} (when the $j=2$ harmonics were exactly conserved) is not quite of the above form, because in that model, there were no odd conserved quantities that coupled via streaming to non-conserved quantities.  In this case, one simply deletes the middle row/column of (\ref{eq:5x5}), and a careful analysis of the resulting matrix inverse in the slow sector leads to $\mathcal{A} \sim 1/\ell_{\mathrm{ee}}$ in the hydrodynamic limit, as we found in the main text.

\section{Joule Heating}
\label{app:joule}
In this short appendix we explicitly show that entropy production is given by (\ref{eq:WTS}) in the main text.   At the quantum level, the many-body density matrix is  \begin{equation}
\varrho \approx \bigotimes_{x,p} \left(f(x,p)|1_{x,p}\rangle\langle 1_{x,p}| + (1-f(x,p))|0_{x,p}\rangle\langle 0_{x,p}|\right).   \label{eq:varrho}
\end{equation}
where we have employed Fermi statistics in an obvious way.  The ket $|n_{x,p}\rangle$ denotes whether a given `quantum state' has 0 or 1 fermions in it.  This is an approximate formula, and should be interpreted as only valid on distances $x$ and $p$ obeying $\mathrm{\Delta}x\mathrm{\Delta}p\gg \hbar$.   The von Neumann entropy is then given by \begin{equation}
S = -\mathrm{tr}[\varrho\log\varrho] = -\int\mathrm{d}^dx \mathrm{d}^dp \left[f(x,p) \log f(x,p) + (1-f(x,p))\log(1-f(x,p))\right],
\end{equation} 
with the latter equality coming from the explicit expression (\ref{eq:varrho}).   Hence, expanding in small perturbations around equilibrium, the time-dependent (unsourced) Boltzmann equation leads to: \begin{align}
\frac{\mathrm{d}S}{\mathrm{d}t} &= -\int \frac{\mathrm{d}^dx \mathrm{d}^dp}{(2\pi\hbar)^dV_x}  \frac{\partial f}{\partial t} \log \frac{f}{1-f}  \notag \\
&\approx \int \frac{\mathrm{d}^dx \mathrm{d}^dp}{(2\pi\hbar)^dV_x}  \left(-\frac{\partial f_{\mathrm{eq}}}{\partial \epsilon}\right) \left[\mathsf{W}\Phi + \mathsf{L}\Phi \right] \left[ \log \frac{f_{\mathrm{eq}}}{1-f_{\mathrm{eq}}} + \frac{1}{f_{\mathrm{eq}}(1-f_{\mathrm{eq}})} \left(-\frac{\partial f_{\mathrm{eq}}}{\partial \epsilon}\right)\Phi   + \cdots \right].
\end{align}
Using the form of the Fermi Dirac distribution (\ref{eq:fermi}), we can simplify the above expression: \begin{equation}
\frac{\mathrm{d}S}{\mathrm{d}t} = \int \frac{\mathrm{d}^dx \mathrm{d}^dp}{(2\pi\hbar)^dV_x}  \left(-\frac{\partial f_{\mathrm{eq}}}{\partial \epsilon}\right) \left[\mathsf{W}\Phi + \mathsf{L}\Phi\right] \left[ - \frac{\epsilon + V_{\mathrm{imp}}}{T} + \frac{\Phi}{T}   + \cdots \right].
\end{equation} 
To get rid of the first term in the final brackets, we use the fact that $\epsilon(p)$ and $1$ are -- pointwise in $x$ -- null vectors of $\mathsf{W}$ since they correspond to locally conserved quantities, as well as the fact that $\mathsf{L}$ annihilates $\epsilon + V_{\mathrm{imp}}$ by its definition in (\ref{eq:Ldef}).   Because $\mathsf{W}$ is symmetric and $\mathsf{L}$ is antisymmetric, we readily arrive at $T\dot{S} = \langle \Phi | \mathsf{W}|\Phi\rangle$ using the inner product (\ref{eq:innerproduct}).  This leads to (\ref{eq:WTS}) in the main text.  While (\ref{eq:LAmain}) is a sourced, static problem, we continue to interpret (\ref{eq:WTS}) as the entropy production.

\section{Variational Principle in the Ballistic Limit}
\label{app:varbal}

In this appendix we will show how the variational principle (\ref{eq:bound32}) is consistent with a finite resistivity in the limit where $\mathsf{W}\rightarrow 0$.   For simplicity, let us assume that the matrix \begin{equation}
\mathsf{W} =  z,
\end{equation} with $z$ an (infinitesimal) relaxation rate which we will send to zero.  Also for simplicity, we will neglect the constraints, as they are automatically satisfied when we remove the regulator $z$ at the end of the calculation.   We thus look for a `minimum' of \begin{equation}
\mathcal{R}[\Phi] = \frac{\langle \Phi | \mathsf{W}_{\mathrm{o}} + \mathsf{L}^{\mathsf{T}}\mathsf{W}^{-1}_{\mathrm{e}}\mathsf{L}|\Phi\rangle}{\langle \Phi|\mathsf{J}_x\rangle^2}.
\end{equation}

As $z\rightarrow 0$, it is clear that we should look for vectors where $\mathsf{L}|\Phi\rangle \approx 0$.  More precisely, let us look for vectors where $\mathsf{L}|\Phi\rangle \sim z|\Phi\rangle$.  In the conventional basis $|\vec x \vec p\rangle $, we look for solutions to  \begin{equation}
\mathsf{L}|\Phi\rangle =  \left|\vec v\cdot \frac{\partial \Phi}{\partial \vec x} + \vec F \cdot \frac{\partial \Phi}{\partial \vec p}\right\rangle =  z|\Phi\rangle. \label{eq:146}
\end{equation}
Solutions to this differential equation are of the form 
\begin{equation}
|\Phi_{\vec x(0)\vec p(0)}\rangle = \int \mathrm{d}t \; \mathrm{e}^{-z|t|} |\vec x(t)\vec p(t)\rangle. \label{eq:Phixp}
\end{equation}
where \begin{equation}
\frac{\mathrm{d}\vec x}{\mathrm{d}t} = \vec v, \;\;\;\; \frac{\mathrm{d}\vec p}{\mathrm{d}t} = \vec F. \label{eq:newton}
\end{equation}
(\ref{eq:Phixp}) should be interpreted as follows.   We may choose arbitrary initial conditions $\vec x(0)$ and $\vec p(0)$.  By solving (\ref{eq:newton}), we obtain a trajectory $\vec x(t)$ and $\vec p(t)$.  The subscript on $|\Phi_{\vec x(0)\vec p(0)}\rangle$ then simply denotes that we have chosen a solution to (\ref{eq:146}) for the initial conditions specified in the subscript.   At this point, we also note that when $z\rightarrow 0$, we recover solutions to the non-interacting Boltzmann equation, which satisfies current and energy conservation.

We now introduce a set $S$, which intuitively consists of a set of initial conditions $\vec x(0)$ and $\vec p(0)$ which -- after time evolution for a time of order $z^{-1}$, according to (\ref{eq:newton}) -- `cover' the classical phase space.  $S$ is an infinite set.  We will now evaluate $\mathcal{R}[\Phi]$ via a scaling argument on trial functions of the form  \begin{equation}
|\Phi\rangle = \int_{(\vec x(0),\vec p(0))\in S} \mathrm{d}S |\Phi_{\vec x(0)\vec p(0)}\rangle.  \label{eq:Eansatz}
\end{equation}   
For some purposes, it is easiest to think about $|\Phi\rangle$ in the form (\ref{eq:Eansatz}), but for other purposes it will be more instructive to write \begin{equation}
|\Phi\rangle \equiv \int \mathrm{d}^dx \mathrm{d}^p\; \mathrm{e}^{-Z(\vec x, \vec p)} |\vec x \vec p\rangle.
\end{equation} 
Because $S$ is a continuous, the function $Z(\vec x, \vec p)$ is $\mathcal{O}(1)$.

We start by evaluating the numerator of $\mathcal{R}$, which is elementary.  Since $\mathsf{L}|\Phi\rangle \sim \mathsf{W}|\Phi\rangle$ on ans\"atze of the form (\ref{eq:Eansatz}), we conclude that \begin{align}
\langle \Phi | \mathsf{W}_{\mathrm{o}} &+ \mathsf{L}^{\mathsf{T}}\mathsf{W}^{-1}_{\mathrm{e}}\mathsf{L}|\Phi\rangle \sim \langle \Phi|\mathsf{W}_{\mathrm{o}}|\Phi\rangle   \sim z \int \frac{\mathrm{d}^dx \mathrm{d}^dp}{(2\pi\hbar)^d V_x} \left(-\frac{\partial f_{\mathrm{eq}}}{\partial \epsilon}\right) \mathrm{e}^{-2Z(\vec x, \vec p)}  \sim  \nu z.  \label{eq:Enum}
\end{align}
The last step above follows simply from the fact that $Z$ is not parametrically small, by definition of $S$.

The denominator  of $\mathcal{R}$ is a little more subtle.  On a single trajectory, we have \begin{align}
\langle \mathsf{J}_x|\Phi_{\vec x(0)\vec p(0)}\rangle &= -e\int \frac{\mathrm{d}^dx \mathrm{d}^dp}{ (2\pi\hbar)^d V_x  } \left(-\frac{\partial f_{\mathrm{eq}}}{\partial \epsilon}\right)  v_x(\vec x, \vec p) \langle \vec x \vec p|  \int \mathrm{d}t \; \mathrm{e}^{-z|t|} |\vec x(t)\vec p(t)\rangle  \notag \\
&= -\frac{e}{ (2\pi\hbar)^d V_x} \left(-\frac{\partial f_{\mathrm{eq}}}{\partial \epsilon}\right) \int \mathrm{d}t  \; \mathrm{e}^{-z |t|} v_x(t),
\end{align}
where $v_x(t)$ is now the $x$-velocity of the quasiparticle at time $t$, given that it started at position $(\vec x(0), \vec p(0))$;  it can be found by solving (\ref{eq:newton}).  In this formula, we have applied (\ref{eq:Eansatz}) directly, because it is easier to account for how $v_x(t)$ evolves along each trajectory separately.  Under the assumption that (in spatial dimensions $d>1$) the quasiparticles undergo random walks at long times \cite{papanicolaou, lebowitz}, we estimate that on a typical trajectory, the quasiparticle displacement \begin{equation}
\left|\mathrm{\Delta} x\right| \equiv \left|\int \mathrm{d}t  \; \mathrm{e}^{-z |t|} v_x(t) \right| \sim \sqrt{\frac{D}{z}},  \label{eq:displacement}
\end{equation}
where $D$ is a suitable diffusion constant.   In the smooth potential problem that we are studying,  $D\sim v_{\mathrm{F}}\xi$ up to a dimensionless constant, which may be large at weak disorder.

In order to get a strong upper bound on the resistivity, we must maximize the typical size of $\langle E|\Phi\rangle^2$. This leads to the following consideration:  if we include all possible trajectories in $S$, then half of the time $\mathrm{\Delta} x$ is positive, and half of the time it is negative; only the magnitude was given in (\ref{eq:displacement}).  We then expect  the disorder averaged \begin{equation}
\mathbb{E}\left[\langle\mathsf{J}_x|\Phi\rangle^2\right] = \mathbb{E}\left[\left(-e\int \frac{\mathrm{d}S}{ (2\pi\hbar)^d V_x} \;\left(-\frac{\partial f_{\mathrm{eq}}}{\partial \epsilon}\right) \mathrm{\Delta}x(\vec x(0),\vec p(0))\right)^2\right] \rightarrow 0.
\end{equation}
 There are an infinite number of trajectories in $S$,  and while infinitesimally close trajectories (at any finite $z$) will be correlated, we expect that in the thermodynamic limit this average vanishes: trajectories starting at very distant $\vec x(0)$ will be completely uncorrelated.   Also note that in this formula, we integrate over $\vec x$ and $\vec p$ by first integrating over each trajectory (to replace $v_x$ with  $\mathrm{\Delta} x$), and then we integrate over all possible starting points.

We must, therefore, slightly improve our definition of $|\Phi\rangle$.  A simple thing to do is to replace the set $S$ by the set $S^\prime$, consisting only of trajectories where $\mathrm{\Delta} x \ge c \sqrt{D/z}$, with $c>0$ some finite $\mathcal{O}(1)$ number.   This does not change the estimate in (\ref{eq:Enum}).   But now, because all $\mathrm{\Delta}x>0$, we may estimate \begin{equation}
\mathbb{E}\left[\langle \mathsf{J}_x|\Phi\rangle^2\right] = \mathbb{E}\left[\left(-e\int \frac{\mathrm{d}S^\prime}{ (2\pi\hbar)^d V_x} \; \left(-\frac{\partial f_{\mathrm{eq}}}{\partial \epsilon}\right) \mathrm{\Delta}x(\vec x(0),\vec p(0))\right)^2\right] \sim \mathbb{E}\left[\left|\mathrm{\Delta}x\right|^2 \right] (z\nu e)^2 \sim z\nu^2 e^2 D. \label{eq:Edom}
\end{equation}
The second scaling argument in this equation follows from the fact that the $(-\partial f_{\mathrm{eq}}/\partial \epsilon)/ (2\pi\hbar)^dV_x$ weighted integral over the set $S^\prime$ scales as $\nu \times z$ -- the factor of $z$ comes from the fact that we cannot include two points in $S^\prime$, if (\emph{i}) they lie on the same single particle trajectory, and (\emph{ii}) a particle can move from one point to the other in a time $\lesssim z^{-1}$.  In (\ref{eq:Edom}) this extra factor of $z$ has arisen, in contrast to (\ref{eq:Enum}), due to the fact that trajectories are weighted by $v_x$.

Combining (\ref{eq:Enum}) and (\ref{eq:Edom}) we obtain \begin{equation}
\rho \lesssim \mathcal{R}[\Phi] \sim \frac{1}{e^2 \nu D},
\end{equation}
which is the Sommerfeld-Drude scaling for the conductivity of a non-interacting electron gas:  in particular, it will scale as $1/\xi$, as stated in the main text.

\end{appendix}

\bibliographystyle{ourbst}
\addcontentsline{toc}{section}{References}

\bibliography{kinetic}

\end{document}